\documentclass[12pt]{article}%
\usepackage{graphicx}
\usepackage{amsmath}
\usepackage{amsfonts}
\usepackage{amssymb}
\begin{document}

\title{Generalized Equivalence Principle in Extended New General Relativity}

\author{Toshiharu Kawai\footnote{E-mail: kawai@sci.osaka-cu.ac.jp},
Kaoru Shibata and Izumi Tanaka\footnote{E-mail: itanaka@sci.osaka-cu.ac.jp}\\Department of Physics, Osaka City University, \\Osaka 558-8585, Japan}
\maketitle

\begin{abstract}
In extended new general relativity, which is formulated as a reduction of a
$\overline{\mbox{Poincar\'e}}\;$gauge theory of gravity whose gauge group is 
the covering group of the Poincar\'e group, we study the problem of whether the total energy-momentum, total angular momentum and total charge are equal to the corresponding quantities of the gravitational source. We examine this problem for 
charged axi-symmetric solutions of gravitational field equations. Our main concern is the restriction on the asymptotic form of the gravitational field variables imposed by the requirement that physical quantities of the total system are equivalent to the corresponding quantities of the charged rotating source body. This requirement can be regarded as an equivalence principle in a generalized sense.
\end{abstract}

\section{Introduction}
\label{sec:intro}

Energy-momentum, angular momentum and electric charge play central roles in modern physics. The conservation properties of the first two are related to the homogeneity and isotropy of space-time, respectively, and charge conservation corresponds to the invariance of the action integral  
under internal $U(1)$ transformations. Also, local quantities such as energy-momentum density, angular momentum density and charge density are well defined if gravitational fields are not present in the system in question. 

In general relativity (G. R.), however, well-behaved energy-momentum and
angular momentum densities have not yet been defined, although total
energy-momen-\linebreak tum and 
total angular momentum can be defined for an asymptotically flat space-time 
surrounding an isolated finite system. The total energy of the system is 
regarded as the inertial mass multiplied by the square of the velocity 
of light, and there arises the following question: Is the active gravitational 
mass of the isolated system equal to its inertial mass?  
This question regards an aspect of the equivalence principle, and it is usually considered to be  affirmatively answered within G. R.\cite{Misner-Thorne-Wheeler,Shirafuji-Nashed-Kobayashi} However, the \textit{equality} of the active gravitational
mass and the inertial mass is violated for the Schwarzschild
metric when it is expressed with using a certain coordinate 
system.\cite{Bozhkov-Rodrigues}

New general relativity (N. G. R.),\cite{Hayashi-Shirafuji} which is formulated by gauging coordinate translations and is constructed within the Weitzenb\"{o}ck space-time, is a possible alternative to G. R. The most general gravitational Lagrangian density, which is quadratic in the torsion tensor and is invariant under global Lorentz transformations including also space inversion and general coordinate transformations, has three free parameters, $c_{1}\; ,
c_{2}$ and $c_{3}$. Solar system experiments show that $c_{1}$ and $-c_{2}$ 
are very likely to be equal to $-1/(3\kappa )$. 
In Ref.\cite{Shirafuji-Nashed-Hayashi}, Shirafuji, Nashed and Hayashi give, for the case with $c_{1}=-1/(3\kappa )=-c_{2}$, the most general 
spherically symmetric solution, and they clarified the restriction on the behavior of the vierbeins at spatial infinity imposed 
by the requirement that the inertial mass is equal to the active gravitational mass. Such analysis has been extended to the case in which 
$c_{1}\neq -1/(3\kappa )\neq -c_{2}$ in Ref.\cite{Shirafuji-Nashed-Kobayashi}.

Extended new general relativity (E. N. G. R.),\cite{Kawai-Toma} is obtained as a reduction of the $\overline{\mbox{Poincar\'e}}$ gauge theory (\={P}. G. T.) of 
gravity,\cite{Kawai01} which is formulated on the basis of the principal fiber
bundle over the space-time, possessing the covering group $\bar{P}_{0}$ of the
Poincar\'{e} group as the structure group, by following the lines of the
standard geometric formulation of Yang-Mills theories as closely as possible.
E. N. G. R.\cite{Kawai-Toma} is also constructed within the Weitzenb\"{o}ck space-time and has many points in common with N. G. R. The field equations for the vierbeins 
in E. N. G. R. , for example, are identical to those in N. G. R., if fields with non-vanishing \lq \lq intrinsic" energy-momentum do not exist. 

In this paper, considering charged axi-symmetric solutions in E. N. G. R., we  examine the condition imposed on the asymptotic behavior of the field variables by the requirement\footnote{This is a generalization of the 
requirement that the inertial mass is equal to the active gravitational  mass.} that the total energy-momentum, the total angular momentum and the 
total charge of the system are all equal to the corresponding quantities 
of the central gravitating body.

This paper is organized as follows. In \S 2, we give the basic formulation
of E. N. G. R. In \S 3, we calculate conserved quantities for a charged axi-symmetric solution. In \S 4, we study solutions obtainable from the 
solution discussed in \S 3 and examine restrictions on the 
field variables imposed by the requirement mentioned above. In the final section,
we give a summary and discussion.

\section{Basic formulation}
\label{sec:basic}
\subsection{$\overline{\mbox{Poincar\'e}}$ gauge theory}
\label{subsec:PbarGT}
\=P. G. T.\cite{Kawai01} is formulated on the basis of the
principal fiber bundle ${\cal P}$ over the space-time $M$ possessing 
the covering group $\bar{P}_{0}$ of the proper orthochronous Poincar\'{e} group as the structure group. The space-time is assumed to be a noncompact four-
dimensional differentiable manifold having a countable base. The bundle ${\cal P}$ admits a 
connection $\Gamma $. The translational and rotational parts of the coefficients of $\Gamma $ will be written as $A^{k}{}_{\mu }$
and $A^{k}{}_{l\mu }$, respectively. The fundamental field variables
are $A^{k}{}_{\mu }\;,A^{k}{}_{l\mu }$, the Higgs-type field 
$\psi=\{\psi^{k}\}$, and 
the matter field $\phi=\{\phi^{A}|A=1,2,3,\cdots ,N\}$.\footnote{Unless otherwise stated, we use the following conventions for indices: The middle part of the Greek alphabet, $\mu,\nu,\lambda,\cdots$, denotes 0, 1, 2 and 3, while the initial part, $\alpha,\beta,\gamma,\cdots$, denotes 1, 2 and 3. In a similar way, the middle part of the Latin alphabet, $i,j,k,\cdots$, denotes 0, 1, 2 and 3, unless otherwise stated, while the initial part, $a,b,c,\cdots$, denotes 1, 2 and 3. The capital letters $A$ and $B$ are used for indices of components of the field $\phi $, and $N$ denotes the dimension of the representation $\rho$.}
These fields transform as\footnote{For the function $f$ on $M$, we define 
$f_{,\mu}\stackrel{\mbox{\scriptsize def}}{=}\partial f/\partial x^{\mu }$.}
\begin{eqnarray}
\label{eq:transformation}
\psi^{\prime k}&=&(\Lambda(a^{-1}))^{k}{}_{l}(\psi^{l}-t^{l})\; ,\nonumber \\
A^{\prime k}{}_{\mu }&=&
(\Lambda(a^{-1}))^{k}{}_{l}(A^{l}{}_{\mu }+t^{l}{}_{,\mu}+A^{l}{}_{m\mu }t^{m})\; ,
\nonumber \\
A^{\prime k}{}_{l\mu }&=&(\Lambda(a^{-1}))^{k}{}_{m}A^{m}{}_{n\mu }
(\Lambda(a))^{n}{}_{l}
+(\Lambda(a^{-1}))^{k}{}_{m}(\Lambda(a))^{m}{}_{l,\mu}\; ,\nonumber \\
\phi^{\prime A}&=&(\rho((t,a)^{-1}))^{A}{}_{B}\phi^{B}\; ,
\end{eqnarray}
under the $\overline{\mbox{Poincar\'e}}$ gauge transformation
\begin{equation}
\label{eq:P-trans}
\sigma^{\prime }(x)
=\sigma(x)\cdot (t(x),a(x))\; ,\; \; t(x)\in T^{4}\; ,\; \; a(x)\in SL(2,C)\; .
\end{equation}
Here, $\Lambda $ is the covering map from $SL(2,C)$ to the proper orthochronous
Lorentz group, and $\rho $ stands for the representation of the 
$\overline{\mbox{Poincar\'e}}$ group to which the field $\phi $ belongs. Also, $\sigma $ and $\sigma^{\prime }$ represent the local cross sections of 
${\cal P}$. The dual components $b^{k}{}_{\mu}$ of the vierbeins 
$b^{\mu }{}_{k}\partial /\partial x^{\mu }$ are related to the
field $\psi$ and the gauge potentials $A^{k}{}_{\mu }$ and $A^{k}{}_{l\mu }$
through the relation
\begin{equation}
\label{eq:vierbein-comp}
b^{k}{}_{\mu}=\psi^{k}{}_{,\mu }+A^{k}{}_{l\mu}\psi^{l}+A^{k}{}_{\mu}
\; ,
\end{equation}
and these transform according to
\begin{equation}
\label{eq:vierbein-trans}
b^{\prime k}{}_{\mu }=(\Lambda(a^{-1}))^{k}{}_{l}b^{l}{}_{\mu }\; 
\end{equation}
under the transformation (\ref{eq:P-trans}). Also, they are related to the metric 
$g_{\mu \nu }dx^{\mu }\otimes dx^{\nu }$ of $M$ through the relation 
\begin{equation}
\label{eq:metric-vierbeins}
g_{\mu \nu }=\eta_{kl}b^{k}{}_{\mu }b^{l}{}_{\nu }\; ,
\end{equation}
with $(\eta_{kl})\stackrel{\mbox{\scriptsize def}}{=}{\mbox{diag}}(-1,1,1,1)$.

There is a 2 to 1 bundle homomorphism $F$ from ${\cal P}$ to the affine frame bundle
${\cal A}(M)$ over $M$, and an extended spinor structure and a spinor structure
exist in association with it.\cite{Kawai03} The space-time $M$ is orientable, which follows from its assumed noncompactness and from the fact that $M$ has a spinor structure.   

The affine frame bundle ${\cal A}(M)$ admits a connection $\Gamma_{A}$. 
The $T^{4}$-part, $\Gamma^{\mu }{}_{\nu }$, and $GL(4,R)$-part, 
$\Gamma^{\mu }{}_{\nu \lambda }$, of its connection coefficients are related to 
$A^{k}{}_{l\mu }$ and $b^{k}{}_{\mu }$ through the relations
\begin{equation}
\label{eq:connection}
\Gamma^{\mu }{}_{\nu}=\delta^{\mu}{}_{\nu}\; ,\; \; \; 
A^{k}{}_{l\mu}=b^{k}{}_{\nu }b^{\lambda}{}_{l}\Gamma^{\nu }{}_{\lambda \mu }
+b^{k}{}_{\nu }b^{\nu }{}_{l,\mu }\; 
\end{equation}
by the requirement that $F$ maps the connection $\Gamma $ into $\Gamma_{A}$ 
, and the space-time $M$ is of the Riemann-Cartan type.  

The field strengths $R^{k}{}_{l\mu \nu }$, 
$R^{k}{}_{\mu\nu}$ and $T^{k}{}_{\mu \nu }$ of $A^{k}{}_{l\mu }$,
$A^{k}{}_{\mu}$ and of $b^{k}{}_{\mu}$ are given by\footnote{We define 
\begin{eqnarray*}
A_{\cdots [\mu \cdots \nu]\cdots }&\stackrel{\mbox{\scriptsize def}}{=}&
\frac{1}{2}
(A_{\cdots \mu \cdots \nu \cdots }-A_{\cdots \nu \cdots \mu \cdots })\; ,\\
A_{\cdots (\mu \cdots \nu)\cdots }
&\stackrel{\mbox{\scriptsize def}}{=}&\frac{1}{2}
(A_{\cdots \mu \cdots \nu \cdots }+A_{\cdots \nu \cdots \mu \cdots })\;.
\end{eqnarray*}
}
\begin{eqnarray}
\label{eq:field-strength}
R^{k}{}_{l\mu \nu }&\stackrel{\mbox{\scriptsize def}}{=}&
2(A^{k}{}_{l[\nu ,\mu]}+A^{k}{}_{m[\mu}A^{m}{}_{l\nu]})\; ,\nonumber \\
R^{k}{}_{\mu \nu }&\stackrel{\mbox{\scriptsize def}}{=}&
2(A^{k}{}_{[\nu ,\mu]}+A^{k}{}_{l[\mu}A^{l}{}_{\nu]})\; ,\nonumber \\ 
T^{k}{}_{\mu\nu}&\stackrel{\mbox{\scriptsize def}}{=}&
2(b^{k}{}_{[\nu ,\mu ]}+A^{k}{}_{l[\mu}b^{l}{}_{\nu]})\; ,
\end{eqnarray}
and we have the relation
\begin{equation}
\label{eq:field-strength-relation}
T^{k}{}_{\mu\nu}=R^{k}{}_{\mu\nu}+R^{k}{}_{l\mu\nu}\psi^{l}
\; .
\end{equation}
The field strengths $T^{k}{}_{\mu\nu}$ and $R^{kl}{}_{\mu\nu}$ are both
invariant under \emph{internal\/} translations.
The torsion is given by 
\begin{equation}
\label{eq:torsion}
T^{\lambda}{}_{\mu \nu }=2\Gamma^{\lambda}{}_{[\nu \mu]}\; ,
\end{equation}
and the $T^{4}$- and $GL(4,R)$-parts of the curvature are given by
\begin{eqnarray}
\label{eq:curvature}
R^{\lambda}{}_{\mu \nu}&=&
2(\Gamma^{\lambda }{}_{[\nu ,\mu]}+\Gamma^{\lambda }{}_{\rho[\mu }
\Gamma^{\rho}{}_{\nu]})\; ,\\
R^{\lambda }{}_{\rho \mu \nu}&=&
2(\Gamma^{\lambda }{}_{\rho [\nu ,\mu]}+\Gamma^{\lambda }{}_{\tau [\mu }
\Gamma^{\tau}{}_{\rho \nu]})\; ,
\end{eqnarray}
respectively.
Also, we have 
\begin{eqnarray}
\label{eq:strength-curvature-torsion}
T^{k}{}_{\mu \nu}&=&b^{k}{}_{\lambda }T^{\lambda }{}_{\mu \nu }=
b^{k}{}_{\lambda }R^{\lambda}{}_{\mu \nu }\; ,\\
R^{k}{}_{l\mu \nu }&=&b^{k}{}_{\lambda }b^{\rho}{}_{l}
R^{\lambda}{}_{\rho \mu \nu }\; ,
\end{eqnarray}
which follow from Eq. (\ref{eq:connection}).

The covariant derivative of the matter field $\phi $ takes the form
\begin{eqnarray}
\label{eq:cov-derivative matter fields}
D_{k}{\phi}^{A}&=&b^{\mu }{}_{k}D_{\mu }{\phi }^{A}\; ,\nonumber \\
D_{\mu }\phi^{A}&\stackrel{\mbox{\scriptsize def}}{=}&
\partial_{\mu}{\phi}^{A}+\frac{i}
{2}A^{lm}{}_{\mu}(M_{lm}{\phi})^{A}+iA^{l}{}_{\mu}(P_{l}{\phi})^{A}
\; .
\end{eqnarray}
Here, $M_{kl}$ and $P_{k}$ are representation matrices of the standard 
basis of the group $\bar{P}_{0}$ : 
$M_{kl}=-i\rho_{*}(\overline{M}_{kl}),
P_{k}=
-i\rho_{*}(\overline{P}_{k})$. The matrix $P_{k}$ represents the intrinsic energy-momentum of the field $\phi^{A}$,\cite{Kawai03} and it is vanishing for all the observed fields.\footnote{In what follows, the field components 
$b^{k}{}_{\mu }$
and $b^{\mu }{}_{k}$ are used to convert Latin and Greek indices,
in analogy to the case of $D_{k}\phi^{A}$ and $D_{\mu }\phi^{A}$.
Also, raising and lowering the indices $k,l,m,\cdots $ are accomplished with
the aid of $(\eta^{kl})\stackrel{\mbox{\scriptsize def}}{=}
(\eta_{kl})^{-1}$ and $(\eta_{kl})$.}

From the requirement of invariance of the action integral under internal 
$\bar{P}_{0}$ gauge transformations, it follows that the gravitational Lagrangian density is a function of $T_{klm}$ and of $R_{klmn}$. The gravitational Lagrangian density is identical to that in Poincar\'{e} gauge theory, and 
hence \emph{gravitational field equations take the same forms in these theories}.
The field equation for the field $\psi^{k}$ is automatically satisfied if those for $A^{k}{}_{\mu}$ and ${\phi}^{A}$ are both satisfied, and $\psi^{k}$ is a non-dynamical field in this sense.
\subsection{Extended new general relativity}
\label{subsec:ENGR}
\subsubsection{Reduction of $\overline{\mbox{Poincar\'e}}$~ gauge
theory to an extended new general relativity}
\label{subsubsec:reduction}

In \={P}. G. T., we consider the case in which the field strength $R^{kl}{}_{\mu \nu }$ vanishes identically,
\begin{equation}
\label{eq:field-strength-vanish}
R^{kl}{}_{\mu \nu}\equiv0 \; ,
\end{equation}
and we choose $SL(2.C)$-gauge such that
\begin{equation}
\label{eq:gauge-condition}
A^{kl}{}_{\mu}\equiv 0\; .
\end{equation}
Then, the curvature vanishes, and the space-time reduces to the Weitzenb\"{o}ck space-time, which means that we have a teleparallel theory.

Also, the vierbeins $b^{k}{}_{\mu}$, the affine connection coefficients
${\Gamma}^{\lambda}{}_{\mu \nu}$ and the torsion tensor $T^{\lambda}{}_{\mu \nu}$
are given by 
\begin{eqnarray}
\label{eq:vierbein-ENGR}
b^{k}{}_{\mu}&=&{\psi}^{k}{}_{,\mu}+A^{k}{}_{\mu}\; ,\\
\label{eq:affine-connec}
{\Gamma}^{\lambda}{}_{\mu\nu}&=&b^{\lambda}{}_{l}\,b^{l}{}_{\mu,\nu}\; ,
\end{eqnarray}
and
\begin{equation}
\label{eq:torsion-strength}
T^{\lambda}{}_{\mu\nu}=2\,{\Gamma}^{\lambda}{}_{[\nu\mu]}=
2b^{\lambda}{}_{k}b^{k}{}_{[\nu,\mu]}=b^{\lambda}{}_{k}T^{k}{}_{\mu \nu }\; , 
\end{equation}
respectively.
Introducing the volume element $dv$ by 
\begin{equation}
\label{eq:vulume-element}
dv\stackrel{\mbox{\scriptsize def}}{=}
\sqrt{-g}dx^{0}\wedge dx^{1}\wedge dx^{2}\wedge dx^{3}\; ,
\end{equation}
with $g\stackrel{\mbox{\scriptsize def}}{=}\det(g_{\mu \nu})=
-\{\det(b^{k}{}_{\mu })\}^{2}$, we consider the action integral 
\begin{equation}
\label{eq:action}
\mbox{\boldmath{$I$ }}=\int_{\cal D}L(\psi^{k}{}_{,\mu }, \psi^{k}, A^{k}{}_{\mu,\nu},A^{k}{}_{\mu },A_{\mu ,\nu},A_{\mu }, \phi^{A}{}_{,\mu },\phi^{A}
,\phi^{*A}{}_{,\mu },\phi^{*A})dv\; ,  
\end{equation}
where $A_{\mu }$ and ${\cal D}$ denote the electromagnetic 
vector potential and a compact region in $M$, 
respectively.\footnote{We consider the case in which 
the electromagnetic field and the charged field are present. The field $\phi^{A}$ 
is considered to be a field with the electric charge $q$. This is preparation for the 
subsequent sections, in which space-time produced by a charged source is treated.} 
Also, the symbol $*$ represents the operation of complex
conjugation, and thus $\phi^{*A}$ denotes the complex conjugate of 
$\phi^{A}$.

We impose the following requirement:\\ 
(R.i) The action $\mbox{\boldmath$I$}$ is invariant under {\em local}
internal translations and {\em global } $SL(2,C)$-transformations. 
From this, the identities \cite{Kawai-Toma}
\begin{eqnarray}
\label{eq:identity-1}
& &\frac{\delta \mbox{\boldmath$L$}}{\delta \psi^{k}}
+\partial_{\mu }\left(\frac{\delta \mbox{\boldmath$L$}}{\delta A^{k}{}_{\mu }}\right)+
i\frac{\delta \mbox{\boldmath$L$}}{\delta \phi^{A}}(P_{k}\phi )^{A}
-i\frac{\delta \mbox{\boldmath$L$}}{\delta \phi^{*A}}(P_{k}\phi )^{*A}\equiv 0\;, \\
\label{eq:identity-2}
& &\mbox{\boldmath$F$}_{k}{}^{(\mu \nu)}\equiv 0\; ,\\
\label{eq:identity-3}
& &{}^{\mbox{\rm{\footnotesize tot}}}{\mbox{\boldmath$T$}}_{k}{}^{\mu }
-\partial_{\nu }{\mbox{\boldmath$F$}}_{k}{}^{\mu \nu }
-\frac{\delta \mbox{\boldmath$L$}}{\delta A^{k}{}_{\mu }}\equiv 0\; ,\\
\label{eq:identity-4}
& &\partial_{\mu }{}^{\mbox{\rm{\footnotesize tot}}}{\mbox{\boldmath$S$}}_{kl}{}^{\mu }
-2\frac{\delta \mbox{\boldmath$L$}}{\delta \psi^{[k}}\psi_{l]}
-2\frac{\delta \mbox{\boldmath$L$}}{\delta A^{[k}{}_{\mu }}A_{l]\mu}
-i\frac{\delta \mbox{\boldmath$L$}}{\delta \phi^{A}}(M_{kl}\phi)^{A}
+i\frac{\delta \mbox{\boldmath$L$}}{\delta \phi^{*A}}(M_{kl}\phi)^{*A}\equiv 0\; 
\nonumber \\
& &{}
\end{eqnarray}
follow, where we have defined 
\begin{eqnarray}
\label{eq:def-1}
\mbox{\boldmath$L$}&\stackrel{\mbox{\scriptsize def}}{=}&\sqrt{-g}L\; ,\\
\label{eq:def-2}
\mbox{\boldmath$F$}_{k}{}^{\mu \nu }&\stackrel{\mbox{\scriptsize def}}{=}&
\frac{\partial \mbox{\boldmath$L$}}{\partial A^{k}{}_{\mu,\nu }}\; ,\\
\label{eq:def-3}
{}^{\mbox{\rm \footnotesize{tot}}}{\mbox{\boldmath$T$}}_{k}{}^{\mu }
&\stackrel{\mbox{\scriptsize def}}{=}&
\frac{\partial {\mbox{\boldmath$L$}}}{\partial \psi^{k}{}_{,\mu }}
+i\frac{\partial \mbox{\boldmath$L$}}{\partial \phi^{A}{}_{,\mu }}(P_{k}\phi)^{A}
-i\frac{\partial \mbox{\boldmath$L$}}{\partial \phi^{*A}{}_{,\mu
}}(P_{k}\phi)^{*A}\; ,\\
\label{eq:def-4}
{}^{\mbox{\rm \footnotesize tot}}{{\mbox{\boldmath$S$}}}_{kl}{}^{\mu }
&\stackrel{\mbox{\scriptsize def}}{=}&
-2\frac{\partial \mbox{\boldmath$L$}}{\partial \psi^{[k}{}_{,\mu }}\psi_{l]}
-2{\mbox{\boldmath$F$}}_{[k}{}^{\nu \mu }A_{l]\nu }\nonumber \\
& &-i\frac{\partial \mbox{\boldmath$L$}}{\partial \phi^{A}{}_{,\mu }}(M_{kl}\phi)^{A}
+i\frac{\partial {\mbox{\boldmath$L$}}}{\partial \phi^{*A}{}_{,\mu }}(M_{kl}\phi)^{*A}\; .
\end{eqnarray}
By virtue of the identity (\ref{eq:identity-3}), 
the density ${}^{\mbox{\rm \footnotesize{tot}}}{\mbox{\boldmath$T$}}_{k}{}^{\mu }$ can be expressed in the usual form of the currrent in Yang-Mills theories,
\begin{equation}
\label{eq:dyn-EM-density-alt}
{}^{\mbox{\rm \footnotesize{tot}}}{\mbox{\boldmath$T$}}_{k}{}^{\mu }
=\frac{\partial \mbox{\boldmath$L$}}{\partial A^{k}{}_{\mu }}\;.
\end{equation}

When the field equations $\delta \mbox{\boldmath$L$}/\delta A^{k}{}_{\mu }
\stackrel{\mbox{\scriptsize def}}{=}
\partial{\mbox{\boldmath$L$}}/\partial A^{k}{}_{\mu}
-\partial_{\nu }(\partial{\mbox{\boldmath$L$}}/\partial A^{k}{}_{\mu ,\nu})=0$ and
$\delta {\mbox{\boldmath$L$}}/\delta \phi^{A}=0$ are both satisfied, 
we have the following:\\
(i) The field equation $\delta {\mbox{\boldmath$L$}}/{\delta \psi^{k}}=0$ 
is automatically satisfied, and hence $\psi^{k}$ is not an independent dynamical variable.\\
(ii) 
\begin{eqnarray}
\label{eq:diff-consv-1}
\partial_{\mu }{}^{\mbox{\rm \footnotesize{tot}}}{\mbox{\boldmath$T$}}_{k}{}^{\mu }=0\; ,\\
\label{eq:diff-consv-2}
\partial_{\mu }{}^{\mbox{\rm \footnotesize{tot}}}{\mbox{\boldmath$S$}}_{kl}{}^{\mu }=0\; ,
\end{eqnarray}
which are the differential conservation laws of the dynamical
energy-momentum and the ``spin'' angular momentum, respectively. The
assertions (i) and (ii) follow from Eqs. (\ref{eq:identity-1}) -- (\ref{eq:identity-4}).

Also, we require the following:\\
(R.ii) The Lagrangian density $L$ is a scalar field on $M$.\\
Then, we have 
\begin{equation}
\label{eq:identity-5}
\tilde{\mbox{\boldmath$T$}}_{\mu }{}^{\nu }-\partial_{\lambda }
{\mbox{\boldmath$\Psi $}}_{\mu }{}^{\nu \lambda }
-\frac{\delta {\mbox{\boldmath$L$}}}{\delta A^{k}{}_{\nu }}
A^{k}{}_{\mu }-\frac{\delta {\mbox{\boldmath$L$}}}{\delta A_{\nu }}A_{\mu }
\equiv 0
\end{equation}
and 
\begin{equation}
\label{eq:identity-6}
{\mbox{\boldmath$\Psi $}}_{\lambda}{}^{(\mu \nu )}\equiv 0\; ,
\end{equation}
with 
\begin{equation}
\label{eq:def-5}
\tilde{\mbox{\boldmath$T$}}_{\mu }{}^{\nu }\stackrel{\mbox{\scriptsize def}}{=}
\delta_{\mu }{}^{\nu }{\mbox{\boldmath$L$}}-{\mbox{\boldmath$F$}}_{k}{}^{\lambda \nu }
A^{k}{}_{\lambda ,\mu }
-{\mbox{\boldmath$F$}}^{\lambda \nu }A_{\lambda, \mu }
-\frac{\partial {\mbox{\boldmath$L$}}}{\partial \phi^{A}{}_{,\nu }}\phi^{A}{}_{,\mu }
-\frac{\partial {\mbox{\boldmath$L$}}}{\partial \phi^{*A}{}_{,\nu }}\phi^{*A}{}_{,\mu }
-\frac{\partial {\mbox{\boldmath$L$}}}{\partial \psi^{k}{}_{,\nu }}\psi^{k}{}_{,\mu }
\end{equation}
and 
\begin{eqnarray}
\label{eq:def-6}
{\mbox{\boldmath$\Psi $}}_{\lambda}{}^{\mu \nu }
&\stackrel{\mbox{\scriptsize def}}{=}&
{\mbox{\boldmath$F$}}_{k}{}^{\mu \nu }A^{k}{}_{\lambda}+
{\mbox{\boldmath$F$}}^{\mu \nu }A_{\lambda }\; , \\
\label{eq:def-7}
{\mbox{\boldmath$F$}}^{\mu \nu }&\stackrel{\mbox{\scriptsize def}}{=}&
\frac{\partial {\mbox{\boldmath$L$}}}{\partial A_{\mu ,\nu }}\; .
\end{eqnarray}

The identities (\ref{eq:identity-5}) and (\ref{eq:identity-6}) 
lead to 
\begin{eqnarray}
\label{eq:diff-consv-3}
\partial_{\nu }\tilde{\mbox{\boldmath$T$}}_{\mu }{}^{\nu }&=&0\; ,\\
\label{eq:diff-consv-4}
\partial_{\nu}{\mbox{\boldmath$M$}}_{\lambda}{}^{\mu \nu }&=&0\; 
\end{eqnarray}
when $\delta {\mbox{\boldmath$L$}}/\delta A^{k}{}_{\mu }=0\; {\rm and}~~ 
\delta {\mbox{\boldmath$L$}}/\delta A_{\mu }=0$, where 
${\mbox{\boldmath$M$}}_{\lambda }{}^{\mu \nu }\stackrel{\mbox{\scriptsize def}}{=}2({\mbox{\boldmath$\Psi $}}_{\lambda }{}^{\mu \nu }-x^{\mu }
\tilde{\mbox{\boldmath$T$}}_{\lambda}{}^{\nu })$. Equations (\ref{eq:diff-consv-3}) and (\ref{eq:diff-consv-4}) are the differential conservation laws of the canonical energy-momentum and ``extended orbital angular momentum'',\cite{Kawai-Saitoh} respectively. We also require invariance of the action under the
$U(1)$ gauge transformation:
\begin{equation}
\label{eq:u1-gauge}
A^{\prime }{}_{\mu }=A_{\mu }+\lambda_{,\mu }\; ,\; \; \; 
\phi^{\prime A}=\exp (iq\lambda)\phi^{A}\; ,\; \; \; 
\psi^{\prime k}=\psi^{k}\; ,\; \; \; 
A^{\prime k}{}_{\mu }=A^{k}{}_{\mu },
\end{equation}
with $\lambda $ being an arbitrary function on $M$, from which we can obtain
\begin{eqnarray}
\label{eq:identity-7}
& &{\mbox{\boldmath$F$}}^{(\mu \nu )}\equiv 0\; ,\\
\label{eq:identity-8}
& &{\mbox{\boldmath$J$}}^{\mu }
+iq\left(\frac{\partial \mbox{\boldmath$L$}}{\partial \phi^{A}{}_{,\mu }}\phi^{A}
-\frac{\partial {\mbox{\boldmath$L$}}}{\partial \phi^{*A}{}_{,\mu }}\phi^{*A}\right)
\equiv 0\; ,\\
\label{eq:identity-9}
& &\partial_{\mu }\left(\frac{\delta {\mbox{\boldmath$L$}}}{\delta A_{\mu }}
-{\mbox{\boldmath$J$}}^{\mu }\right)\equiv 0\; ,
\end{eqnarray}
with 
\begin{equation}
\label{eq:def-8}
{\mbox{\boldmath$J$}}^{\mu }\stackrel{\mbox{\scriptsize def}}{=}
\frac{\partial {\mbox{\boldmath$L$}}}{\partial A_{\mu }}\; .
\end{equation}
From the identity (\ref{eq:identity-9}), the differential conservation law of the electric charge,
\begin{equation}
\label{eq:diff-consv-5}
\partial_{\mu }{\mbox{\boldmath$J$}}^{\mu }=0\; ,
\end{equation}
follows when the field equation $\delta {\mbox{\boldmath$L$}}/\delta A_{\mu }=0$ 
is satisfied. 
The functional dependence of $L$ is restricted as 
\begin{equation}
\label{eq:func-Lag}
L={\cal L}(\psi^{k}, T_{klm},F_{\mu \nu},\nabla_{k}\phi^{A},\phi^{A},
\nabla^{*}{}_{k}\phi^{*A},\phi^{*A})\; ,
\end{equation}
with ${\cal L}$ satisfying certain identities,\cite{Kawai-Toma}
where 
\begin{eqnarray}
\label{eq:em-F}
F_{\mu \nu }&\stackrel{\mbox{\scriptsize def}}{=}&
\partial_{\mu }A_{\nu }-\partial_{\nu }A_{\mu}\; ,\\
\label{eq:cov-der}
\nabla_{k}\phi^{A}
&\stackrel{\mbox{\scriptsize def}}{=}&
b^{\mu }{}_{k}\nabla_{\mu }\phi^{A}
\stackrel{\mbox{\scriptsize def}}{=}
b^{\mu }{}_{k}(\phi^{A}{}_{,\mu }+
iA^{k}{}_{\mu }(P_{k}\phi)^{A}-iqA_{\mu }\phi^{A})\; ,\nonumber \\
\nabla^{*}{}_{k}\phi^{*A}&\stackrel{\mbox{\scriptsize def}}{=}&
b^{\mu }{}_{k}\nabla^{*}{}_{\mu }\phi^{*A}
\stackrel{\mbox{\scriptsize def}}{=}
b^{\mu }{}_{k}(\phi^{*A}{}_{,\mu }-
iA^{k}{}_{\mu }(P_{k}\phi)^{*A}+iqA_{\mu }\phi^{*A})\; .
\end{eqnarray}

The gravitational action\footnote{This action is identical to the gravitational action in N. G. R.\cite{Hayashi-Shirafuji}} 
\begin{equation}
\label{eq:gravitational-action}
\tilde{\mbox{\boldmath$I$}}_{G}=\int_{\mathcal{D}}
(c_{1}\,t^{klm}t_{klm}+c_{2}\,v^{k}v_{k}+c_{3}\,a^{k}a_{k})dv\; ,
\end{equation}
with $c_{i}\; (i=1,2,3)$ being real constants, satisfies the requirements (R.i) and (R.ii). 
Here $t_{klm}$, $v_{k}$ and $a_{k}$ are irreducible components of
the torsion tensor:
\begin{eqnarray}
\label{eq:irreducible-comp}
t_{klm}&\stackrel{\mbox{\scriptsize def}}{=}&
\frac{1}{2}(T_{klm}+T_{lkm})+\frac{1}{6}
(\eta_{mk}v_{l}+\eta_{ml}v_{k})-\frac{1}{3}\eta_{kl}v_{m}
\; ,\\
v_{k}&\stackrel{\mbox{\scriptsize def}}{=}&T^{l}{}_{lk}\; ,\\
a_{k}&\stackrel{\mbox{\scriptsize def}}{=}&\frac{1}{6}\varepsilon_{klmn}T^{lmn}
\; ,
\end{eqnarray}
where $\varepsilon_{klmn}$ is a completely anti-symmetric Lorentz tensor
with $\varepsilon_{(0)(1)(2)(3)}=-1$.\footnote{Latin indices are put in parentheses to discriminate them from Greek indices.}

The action $\tilde{\mbox{\boldmath$I$}}_{G}$ with\footnote{We will use natural units in which $\hbar =c=1$.}
\begin{equation}
\label{eq:parameter-choice}
c_{1}=-c_{2}=-\frac{1}{3\kappa }\; ,
\end{equation}
with $\kappa $ being the Einstein gravitational constant agrees quite well with  experimental results.\cite{Hayashi-Shirafuji} In what follows, we assume that the condition
(\ref{eq:parameter-choice}) is satisfied. Thus, our gravitational 
action is 
\begin{equation}
\label{eq:gravitational-action-alt-1}
{\mbox{\boldmath$I$}}_{G}\stackrel{\mbox{\scriptsize def}}{=}
\int_{\mathcal D}L_{G}dv\; , 
\end{equation}
with 
\begin{equation}
\label{eq:g-Lagrangian}
L_{G}\stackrel{\mbox{\scriptsize def}}{=}
-\frac{1}{3\kappa}(t^{klm}t_{klm}-v^{k}v_{k})+c_{3}\,a^{k}a_{k}\; .
\end{equation}
\subsubsection{The gravitational and electromagnetic field equations in vacuum}
\label{subsubsec:field-eq}

The electromagnetic Lagrangian density $L_{em}$ is given by\footnote{Here we use Heaviside-Lorentz rationalized unit.}
\begin{equation}
\label{eq:e.m.-Lagrangian}
L_{em}=-\frac{1}{4}g^{\mu\hspace{0.01pt}\rho}g^{\nu\sigma}F_{\mu
\hspace{0.01pt}\nu}F_{\rho\sigma}\; .
\end{equation}
We consider a system described by the Lagrangian density $L\stackrel{\mbox{\scriptsize def}}{=}L_{G}+L_{em}$. The gravitational and 
electromagnetic field equations for this system are the following:
\begin{eqnarray}
\label{eq:G-eq}
G^{\mu\nu}(\{\})+K^{\mu\nu}&=&\kappa T_{em}^{\mu\nu}\; ,\\
\label{eq:G-eq2}
\partial_{\mu}(\sqrt{-g}J^{kl\mu})&=&0\; ,\\
\label{eq:EM-eq}
{\partial}_{\nu}\left(\sqrt{-g}F^{\mu \nu}\right)&=&0\; .
\end{eqnarray}
Here we have defined
\begin{equation}
\label{eq:Eistein-tensor}
G_{\mu\nu}(\{\})\stackrel{\mbox{\scriptsize def}}{=}
R_{\mu\nu}(\{\})-\frac{1}{2}g_{\mu\nu}R(\{\})\; ,
\end{equation}
and
\begin{equation}
\label{eq:Ricci-tensor}
R_{\mu\nu}(\{\})\stackrel{\mbox{\scriptsize def}}{=}
R^{\lambda}{}_{\mu\lambda\nu}(\{\})\; ,\; \; \; R(\{\})
\stackrel{\mbox{\scriptsize def}}{=}g^{\mu\nu}R_{\mu\nu}(\{\})\; ,
\end{equation}
with the Riemann-Christoffel curvature tensor
\begin{equation}
\label{eq:RC-curv}
R^{\lambda}{}_{\rho\mu\nu}(\{\})\stackrel{\mbox{\scriptsize def}}{=}
2\left(\partial_{[\mu}{\lambda\brace\rho~\nu]}
+{\lambda\brace\sigma~[\mu}{\sigma\brace\rho~\nu]}\right)\; .
\end{equation}
Also, $T_{em}^{\mu\nu}$ is the energy-momentum tensor of the electromagnetic 
field,
\begin{equation}
\label{eq:energy-momentum tensor}
T_{em}^{\mu\nu}\stackrel{\mbox{\scriptsize def}}{=}
F^{\mu\hspace{0.01pt}\rho}F^{\nu\sigma}g_{\rho\sigma}
+g^{\mu\hspace{0.01pt}\nu}L_{em}\; ,
\end{equation}
and the tensors $K^{\mu\nu}$ and $J^{kl\mu}$ are defined by
\begin{equation}
\label{eq:k-tensor}
K^{\mu\nu}\stackrel{\mbox{\scriptsize def}}{=}\frac{\kappa}{{\lambda}}\left[\frac{1}{2}\{\;\varepsilon{}^{\mu\rho\sigma{\lambda}}(T^{\nu}{}_{\rho\sigma}-T_{\rho\sigma}{}^{\nu})+\varepsilon{}^{\nu\rho\sigma{\lambda}}(T^{\mu}
{}_{\rho\sigma}-T_{\rho\sigma}{}^{\mu})\;\}a_{{\lambda}}-\frac{3}{2}a^{\mu
}a^{\nu}-\frac{3}{4}g^{\mu\nu}a^{{\lambda}}a_{{\lambda}}\right]
\end{equation}
and 
\begin{equation}
\label{eq:J-tensor}
J^{kl\mu}\stackrel{\mbox{\scriptsize def}}{=}-\frac{3}{2}b^{k}{}_{\rho}b^{l}
{}_{\sigma}\varepsilon{}^{\rho\sigma\mu\nu}a_{\nu}\; ,
\end{equation}
respectively, where we have used
\begin{equation}
\label{eq:lamda}
{\lambda}\stackrel{\mbox{\scriptsize def}}{=}
\frac{9\kappa}{4\kappa c_{3}+3}\; .
\end{equation}
\subsubsection{An exact solution of gravitational and electromagnetic
field equations with a charged rotating source}
\label{subsubsec:solution}

An exact solution of the field equations (\ref{eq:G-eq})
-- (\ref{eq:EM-eq}), which represents the gravitational and electromagnetic fields surrounding a charged rotating source, is given in 
Ref.\cite{Kawai-Toma02}. This will be described below.
The vector fields $b^{k}{}_{\mu }$ and the electromagnetic potential 
$A_{\mu }$ have the expressions
\begin{eqnarray}
\label{eq:v-fields}
b^{k}{}_{\mu}&=&{}^{(0)}b^{k}{}_{\mu }+\frac{a}{2}l^{k}l_{\mu}-\frac{Q^{2}}
{2}m^{k}m_{\mu}\; ,\\
\label{eq:em-potential}
A_{\mu}&=&-\frac{q}{4\pi}\sqrt{Z}\,l_{\mu}\; ,
\end{eqnarray}
where ${}^{(0)}b^{k}{}_{\mu }$ are the dual components of constant vierbeins  
and are defined by ${}^{(0)}b^{k}{}_{\mu }\stackrel{\mbox{\scriptsize def}}{=}
{\delta}^{k}{}_{\mu}$. The functions $Z ,l_{\mu }, m_{\mu }, l^{k}$ and 
$m^{k}$ are given by 
\begin{eqnarray}
\label{eq:Z-lm}
Z&=&\frac{N}{D}\; ,\nonumber \\
l_{0} &=&\sqrt{Z}\; ,\; \; \; 
l_{\alpha}=\frac{2\sqrt{Z}}{D+r^{2}+h^{2}}\left[Nx_{\alpha}
\;+\frac{h^{2}x^{3}{\delta}^{3}{}_{\alpha}}{N}-{\varepsilon}
_{\alpha\beta3}x^{\beta}\right]\; ,\nonumber \\
m_{\mu }&=&\frac{l_{\mu }}{\sqrt{N}}\; ,\; \; \; 
l^{k}=\delta^{k}{}_{\mu }\eta^{\mu \nu }l_{\nu }\; , \; \; \; 
m^{k}=\delta^{k}{}_{\mu }\eta^{\mu \nu }m_{\nu }\; ,
\end{eqnarray}
where $D$ and $N$ are given by 
\begin{equation}
\label{eq:D-N}
D=\sqrt{(r^{2}-h^{2})^{2}+4h^{2}(x^{3})^{2}}\; ,\; \; \; 
N=\frac{\sqrt{r^{2}-h^{2}+D}}{\sqrt{2}}\; ,
\end{equation}
with $r\stackrel{\mbox{\scriptsize def}}{=}\sqrt{(x^{1})^{2}+(x^{2})^{2}+(x^{3})^{2}}$. In Eq.~(\ref{eq:Z-lm}), the $\varepsilon_{\alpha \beta \gamma}$ are three-dimensional totally anti-symmetric tensor with $\varepsilon_{123}=1$. 
Also, we have defined 
\begin{equation}
\label{eq:a-Q-h}
a\stackrel{\mbox{\scriptsize def}}{=}\frac{\kappa m}{4\pi}\; ,
\quad Q\stackrel{\mbox{\scriptsize def}}{=}\frac{q}{4\pi}\sqrt{\frac{\kappa}{2}
}\;, \; \; \; 
h\stackrel{\mbox{\scriptsize def}}{=}-\frac{J}{m}\; ,
\end{equation}
with $m, J$ and $q$ being the active gravitational mass, the absolute value 
of the angular momentum, and the electromagnetic charge of the source body, respectively. 

For the solution given by Eqs. (\ref{eq:v-fields}) and (\ref{eq:em-potential}), 
the axial vector part $a_{\mu }$ of the torsion tensor vanishes,
\begin{equation}
\label{eq:irre-a}
a_{\mu}=0\; ,
\end{equation}
and the metric is identical to the charged Kerr metric in G. R.

The asymptotic forms of the vierbeins and the electromagnetic vector potential for large $r$ are given by
\begin{eqnarray}
\label{eq:b-asympt}
b^{a}{}_{\alpha}&=&{\delta}^{a}{}_{\alpha}+\frac{1}{2}\left(a-\frac{Q^{2}
}{r}\right)\frac{n^{a}n_{\alpha}}{r}-\frac{ahn^{\beta}}{2r^{2}}\left(
{\varepsilon}_{\alpha\beta3}n^{a}+{\varepsilon}^{a}{}_{\beta3}n_{\alpha
}\right)
+O^{a}{}_{\alpha }\left(\frac{1}{r^{3}}\right)\; ,\nonumber \\
b^{(0)}{}_{\alpha}&=&-\frac{n_{\alpha}}{2r}\left(a-\frac{Q^{2}}{r}\right)
+\frac{ah}{2r^{2}}{\varepsilon}_{\alpha\beta3}n^{\beta}
+O_{\alpha }\left(\frac{1}{r^{3}}\right)\; ,\nonumber \\
b^{a}{}_{0}&=&\frac{n^{a}}{2r}\left(a-\frac{Q^{2}}{r}\right)-
\frac{ah}{2r^{2}}{\varepsilon}^{a}{}_{\beta3}n^{\beta}
+O^{a}\left(\frac{1}{r^{3}}\right)\; ,\nonumber \\
b^{(0)}{}_{0}&=&1-\frac{1}{2r}\left(a-\frac{Q^{2}}{r}\right)
+O\left(\frac{1}{r^{3}}\right)
\end{eqnarray}
and 
\begin{equation}
\label{eq:A-asympt}
A_{0}=-\frac{q}{4\pi r}+O\left(\frac{1}{r^{3}}\right)\; ,\; \; \; 
A_{\alpha}=-\frac{q}{4\pi}\frac{x^{\alpha}}{r^{2}}+O_{\alpha}\left(\frac{1}{r^{3}}\right)\; ,
\end{equation}
respectively, which follow from Eqs.~(\ref{eq:v-fields}) and
(\ref{eq:em-potential}). Here, we have defined \linebreak $n^{\alpha}\stackrel{\mbox{\scriptsize def}}{=}x^{\alpha}/r$, and the expression $O\left(1/r^{w}\right)$ with positive $w$ denotes a term 
such that $\lim_{r\rightarrow\infty}r^{w}O\left(1/r^{w}\right)\,=\,$constant.~\footnote{The symbols $a$ and $\alpha$ in $O^{a}{}_{\alpha }(1/r^{3}), O_{\alpha }(1/r^{3})$ and $O^{a}(1/r^{3})$ are to show that these terms have indices as indicated.}
\subsubsection{Asymptotic form of $\psi^{k}$}
\label{subsubsec:asympt-psi}

The space-time in this theory has vanishing curvature tensor. 
When, additionally,  the torsion tensor vanishes identically, the space-time 
is the Minkowski space-time, for which the
translational gauge potentials $A^{k}{}_{\mu}$ can be chosen to be zero
and Eq. (\ref{eq:vierbein-comp}) is reduced to
\begin{equation}
\label{eq:v-ex}
b^{k}{}_{\mu}={\psi}^{k}{}_{,\mu}\; .
\end{equation}
For the solution given by Eqs. (\ref{eq:v-fields}) and 
(\ref{eq:em-potential}), we have\footnote{The square bracket
$[ \; \;  ]$ in the suffix of $O^{k}{}_{[\mu \nu ]}(1/r^{2})$ 
indicate that this term is anti-symmetric with respect to 
$\mu \; ,\nu $.}  
\begin{equation}
\label{eq:asympt-torsion}
T^{k}{}_{\mu \nu}=O^{k}{}_{[\mu \nu ]}\left(\frac{1}{r^{2}}\right)\; ,
\end{equation}
and the space-time asymptotically approaches the Minkowski 
space-time for large $r$. 

The above discussion and the consideration given in Ref.\cite{Utiyama} to introduce vierbeins suggest that $\psi{}^{k}$ can be regarded as a Minkowskian coordinate at spatial infinity, and we are naturally led to employ the following
form of $\psi{}^{k}$: 
\begin{eqnarray}
\label{eq:psihyou}
{\psi}^{k}&=&{}^{(0)}b^{k}{}_{\mu}x^{\mu}+{}^{(0)}\psi^{k}+\,O^{k}{}_{{}
}\left(\displaystyle\frac{\mathstrut1}{r^{\beta}}\right),\quad
(\beta>0)\; \nonumber \\
{\psi}^{k}{}_{,\mu}&=&{}^{(0)}b^{k}{}_{\mu }+\,O^{k}{}_{\mu}\left(
\displaystyle\frac{\mathstrut1}{r^{1+\beta}}\right)\; ,\nonumber \\
{\psi}^{k}{}_{,\mu\nu}&=&O^{k}{}_{(\mu\nu)}\left(\displaystyle
\frac{\mathstrut1}{r^{2}}\right)\; ,
\end{eqnarray}
where $^{(0)}\psi^{k}$ and $\beta $ are constants.

\section{Equivalence relations for the case of the solution \\
represented by 
Eqs.~(\ref{eq:v-fields}) and (\ref{eq:em-potential})}
\label{sec:equivalence}

In this section, on the basis of the discussion in Refs.\cite{Kawai-Saitoh}
and \cite{Kawai02}, we examine the energy-momentum, the angular momentum and 
the charge for the solution given in the preceding section.

\subsection{The case in which 
$\{{\psi}^{k},A^{k}{}_{\mu},A_{\mu }\}$ is employed as the set of independent 
field variables}
\label{subsec:choice1}

We regard the Lagrangian density ${\mbox{\boldmath$L$}}
\stackrel{\mbox{\scriptsize def}}{=}\sqrt{-g}L$ as a function of ${\psi}^{k},A^{k}{}_{\mu},A_{\mu }$ and of their derivatives. For this case, the generator 
$M_{k}$ of \emph{internal\/} translations and the generator $S_{kl}$ of \emph{internal\/}
Lorentz transformations are\cite{Kawai01} the dynamical energy-momentum and
the total (=spin $+$ orbital) angular momentum, respectively, and they are
expressed as
\begin{equation}
\label{eq:energy-momentum}
M_{k}\stackrel{\mbox{\scriptsize def}}{=}
\int_{\sigma}\!{}^{\mbox{\rm \footnotesize{tot}}}{\mbox{\boldmath$T$}}_{k}{}^{\mu }d{\sigma}_{\mu }
\end{equation}
and
\begin{equation}
\label{eq:angular momentum}
S_{kl}\stackrel{\mbox{\scriptsize def}}{=}
\int_{\sigma}\!{}^{\mbox{\rm \footnotesize{tot}}}{\mbox{\boldmath$S$}}_{kl}{}^{\mu
}d{\sigma}_{\mu}\; .
\end{equation}
Here, $\sigma $ is a space-like surface, and $d\sigma_{\mu }$ is the surface element on it:
\begin{equation}
d\sigma_{\mu }=\frac{1}{3!}\varepsilon_{\mu \nu \lambda \rho }
dx^{\nu }\wedge dx^{\lambda }\wedge dx^{\rho}\; .
\end{equation}
We have 
\begin{equation}
\label{eq:F-Z}
{\mbox{\boldmath$F$}}_{k}{}^{\mu\nu}={\mbox{\boldmath$F$}}^{(1)}{}_{k}{}^{\mu\nu}
+{\mbox{\boldmath$F$}}^{(2)}{}_{k}{}^{\mu\nu}\; ,
\end{equation}
with 
\begin{eqnarray}
\label{eq:spF1}
{\mbox{\boldmath$F$}}^{(1)}{}_{k}{}^{\mu\nu}&\stackrel{\mbox{\scriptsize def}}{=}&
\frac{1}{\kappa\sqrt{-g}}b_{k\rho}\partial{}_{\sigma}\left\{(-g)\,g^{\rho\lbrack\mu}g^{\nu]\sigma}\right\}+{\mbox{\boldmath$Z$}}_{k}{}^{\mu\nu}\; ,\\
\label{eq:spF2}
{\mbox{\boldmath$F$}}^{(2)}{}_{k}{}^{\mu\nu}
&\stackrel{\mbox{\scriptsize def}}{=}&
\frac{3}{2\lambda}\sqrt{-g}\,{\varepsilon}_{k}{}^{lmn}b^{\mu}{}_{m}b^{\nu}{}
_{n}a_{l}\; ,\\
\label{eq:spZ}
{\mbox{\boldmath$Z$}}_{k}{}^{\mu\nu}&\stackrel{\mbox{\scriptsize def}}{=}&
\frac{\sqrt{-g}}{\kappa }\left\{\,b^{[\mu}{}_{k}(\,b^{\nu]l}b^{\lambda}{}_{l.\lambda}-b^{\lambda l}\,b^{\nu]}{}_{l,\lambda})+b^{\lambda }{}_{k}b^{[\mu}{}_{l}
b^{\nu]l}{}_{,\lambda}\right\}\; ,
\end{eqnarray}
as is known by using the explicit form of $L_{G}$.\cite{Kawai-Toma} 
From Eqs. (\ref{eq:vierbein-ENGR}), (\ref{eq:identity-3}), (\ref{eq:def-3}) and 
(\ref{eq:F-Z}), we find that ${}^{\mbox{\rm \footnotesize{tot}}}{\mbox{\boldmath$S$}}_{kl}{}^{\mu }$ defined by 
Eq. (\ref{eq:def-4}) can be rewritten as
\begin{eqnarray}
\label{eq:S-alt}
{}^{\mbox{\rm \footnotesize{tot}}}{\mbox{\boldmath$S$}}_{kl}{}^{\mu }&=&
2\partial_{\nu }(\psi_{[k}
{\mbox{\boldmath$F$}}_{l]}{}^{\mu \nu})
+\frac{2}{\kappa }\partial_{\nu }
\left(\sqrt{-g}b^{\mu }{}_{[k}b^{\nu }{}_{l]}
-{}^{(0)}b^{\mu }{}_{[k}{}^{(0)}b^{\nu }{}_{l]}\right) \nonumber \\
& &-b_{[k\nu}{\mbox{\boldmath$F$}}^{(2)}{}_{l]}{}^{\mu \nu }
+2\psi_{[k}\frac{\delta {\mbox{\boldmath$L$}}}{\delta A^{l]}{}_{\mu }}\; ,
\end{eqnarray}
where ${}^{(0)}b^{\mu }{}_{k}$ are the components of the constant vierbeins:
${}^{(0)}b^{\mu }{}_{k}\stackrel{\mbox{\scriptsize def}}{=}\delta^{\mu }{}_{k}$. 
For the vierbeins given by Eq. (\ref{eq:v-fields}), we have 
\begin{equation}
\label{eq:F-Z-condition}
{\mbox{\boldmath$Z$}}_{k}{}^{\mu \nu}=0\; , \; \; \; 
{\mbox{\boldmath$F$}}^{(2)}{}_{k}{}^{\mu \nu}=0\; ,
\end{equation}
and hence 
\begin{equation}
\label{eq:F02}
{\mbox{\boldmath$F$}}_{k}{}^{\mu \nu}={\mbox{\boldmath$F$}}^{(1)}{}_{k}{}^{\mu \nu}
=\frac{1}{\kappa \sqrt{-g}}b_{k\rho}{\partial}_{\sigma}\left\{(-g)\,g^{\rho
\lbrack\mu}g^{\nu]\sigma}\right\}\; .
\end{equation}
\subsubsection{Energy-momentum}
\label{subsubsec:energy-momentum1}
When the field equation $\delta {\mbox{\boldmath$L$}}/\delta A^{k}{}_{\mu }=0$
is satisfied, Eq. (\ref{eq:energy-momentum}) can be rewritten as
\begin{equation}
\label{eq:energy-momentum-alt}
M_{k}=\int_{\sigma }\partial_{\nu }{\mbox{\boldmath$F$}}_{k}{}^{\mu \nu }d\sigma_{\mu }
=\int_{S}{\mbox{\boldmath$F$}}_{k}{}^{0\alpha }r^{2}n_{\alpha }d\Omega \; ,
\end{equation}
by using the identity (\ref{eq:identity-3}). Here, $S$ and $d\Omega $ stand for the two-dimensional surface of $\sigma$ and the differential solid 
angle, respectively. Equations (\ref{eq:F-Z}), (\ref{eq:F-Z-condition})
-- (\ref{eq:energy-momentum-alt}) and (\ref{eq:asympt-F}) give 
\begin{equation}
\label{eq:energy-momentum-1}
M_{(0)}=-m\; ,\; \; \; M_{a}=0\; .
\end{equation}
The quantity $M_{k}$ is the total energy-momentum vector of the system. 
The first relation in Eq. (\ref{eq:energy-momentum-1}) expresses the \textit{equality} of the active gravitational mass and the inertial mass.
\subsubsection{Angular momentum}
\label{subsubsec:angular-momentum1}
From Eqs. (\ref{eq:angular momentum}) and (\ref{eq:S-alt}), the 
toal angular momentum can be expressed as
\begin{equation}
\label{eq:total-angular-momentum}
S_{kl}=2\int_{\sigma}\!{\partial}_{\nu}\left[\,{\psi}_{[k}
{\mbox{\boldmath$F$}}_{l]}{}^{\mu\nu}+\frac{1}{\kappa}\left\{\sqrt{-g}\,b^{\mu}{}
_{[k}\,b^{\nu}{}_{l]}\,-\,{}^{(0)}b^{\mu}{}_{[k}{}^{(0)}b^{\nu}{}_{l]}\,\right\}
\,\right]d\sigma_{\mu } \; , 
\end{equation}
from which the expression 
\begin{eqnarray}
\label{eq:s-result}
S_{(0)a}&=&{}^{(0)}{\psi}_{a}m=-{}^{(0)}{\psi}_{a}M_{(0)}+{}^{(0)}\psi_{(0)}M_{a}\; ,\nonumber \\
S_{ab}&=&J{\varepsilon}_{ab3}=J{\varepsilon}_{ab3}
+{}^{(0)}{\psi}_{a}M_{b}-{}^{(0)}\psi_{b}M_{a}
\end{eqnarray}
is obtained by the use of Eqs. (\ref{eq:b-asympt}), (\ref{eq:psihyou}),
(\ref{eq:energy-momentum}), (\ref{eq:F-Z}),
(\ref{eq:F02}) and (\ref{eq:asympt-F}).

In the above, terms of the form ${}^{(0)}{\psi}_{k}M_{l}-{}^{(0)}\psi_{l}
M_{k}$ are regarded as to represent the conserved orbital angular momentum around the origin of the internal space.\cite{Kawai02}
Equation (\ref{eq:s-result}) implies that the total angular momentum is equal to the angular momentum of the rotating source.
\subsubsection{Canonical energy-momentum and \lq \lq extended orbital angular momentum"}
\label{subsubsec:canonical1}

The generator $M^{c}{}_{\mu}$ of coordinate translations and the
generator $L_{\mu}{}^{\nu}$ of $GL(4,\mbox{\boldmath$R$})$ coordinate
transformations are the canonical energy-momentum and
the ``extended orbital angular momentum",~\footnote{Note that the
anti-symmetric part $L_{[\mu\nu]}\stackrel{\mbox{\scriptsize def}}{=}
\eta_{[\nu \lambda}L_{\mu ]}{}^{\lambda}$ is the orbital angular momentum.} respectively. They have
the expressions
\begin{eqnarray}
\label{eq:M-L1}
M^{c}{}_{\mu}&\stackrel{\mbox{\scriptsize def}}{=}&
\int_{\sigma}\tilde{\mbox{\boldmath$T$}}_{\mu }{}^{\nu }d{\sigma}_{\nu }
=\int_{\sigma}\!{\partial}_{\tau}{\mbox{\boldmath$\Psi $}}_{\mu}{}^{\nu\tau}d{\sigma}_{\nu}\; ,\\
\label{eq:ack}
L_{\mu}{}^{\nu}&\stackrel{\mbox{\scriptsize def}}{=}&
\int_{\sigma}{\mbox{\boldmath$M$}}_{\mu }{}^{\nu \lambda}d{\sigma}_{\lambda }
=-2\int_{\sigma}\!{\partial}_{\tau}(x^{\nu}{\mbox{\boldmath$\Psi $}}_{\mu}{}^{\lambda\tau})
d{\sigma}_{\lambda}\; .
\end{eqnarray}

The asymptotic behavior of the translational gauge potentials $A^{k}{}_{\mu}$
at spatial infinity is given as
\begin{eqnarray}
\label{eq:asympt-A}
A^{(0)}{}_{0}&=&-\frac{a}{2r}+O\left(\displaystyle
\frac{\mathstrut1}{r^{1+\beta}}\right)\;, \; \; 
A^{(0)}{}_{\alpha}=-\frac{a}{2r}n_{\alpha}+O_{\alpha}
\left(\displaystyle\frac{\mathstrut1}{r^{1+\beta}}\right)\; ,\nonumber \\
A^{a}{}_{0}&=&\frac{a}{2r}n^{a}+O^{a}\left(\displaystyle
\frac{\mathstrut1}{r^{1+\beta}}\right)\; ,\; \; 
A^{a}{}_{\alpha}=\frac{a}{2r}n^{a}n_{\alpha}+O^{a}{}_{\alpha}
\left(\displaystyle\frac{\mathstrut1}{r^{1+\beta}}\right)\; , 
\end{eqnarray}
which are known from Eqs. (\ref{eq:vierbein-ENGR}), (\ref{eq:b-asympt}) and (\ref{eq:psihyou}).
Then $M^{c}{}_{\mu}$ vanishes trivially, 
\begin{equation}
\label{eq:integ-m}
M^{c}{}_{\mu}=
\int_{S}{\mbox{\boldmath$F$}}_{k}{}^{0\alpha}A^{k}{}_{\mu}r^{2}
n_{\alpha }d\Omega=0\; ,
\end{equation}
while $L_{\mu }{}^{\nu}$ is expressed as
\begin{equation}
\label{eq:integ-l}
L_{\mu}{}^{\nu}=-2\int_{S}x^{\nu}
({\mbox{\boldmath$F$}}_{k}{}^{0\alpha}A^{k}{}_{\mu}
+{\mbox{\boldmath$F$}}^{0 \alpha }A_{\mu })r^{2}n_{\alpha}d\Omega\; , 
\end{equation}
and the non-zero components are given by  
\begin{equation}
\label{eq:L-value}
L_{1}{}^{1}=L_{2}{}^{2}=L_{3}{}^{3}=\frac{q^{2}}{6\pi}\; .
\end{equation}
These can be shown by use of Eqs. (\ref{eq:def-6}), (\ref{eq:A-asympt}), (\ref{eq:F-Z}), (\ref{eq:asympt-A}) and (\ref{eq:asympt-F}). 
Thus, the orbital angular momentum $L_{[\mu \nu]}$ is vanishing: 
$L_{[\mu \nu]}=0$.
\subsubsection{Charge}
\label{subsubsec:charge1}
The charge is defined as the generator of $U(1)$ gauge transformations and is given by
\begin{equation}
\label{eq:charge}
C\stackrel{\mbox{\scriptsize def}}{=}\int_{\sigma}\!\partial_{\nu }
\left(\frac{\partial {\mbox{\boldmath$L$}}}{\partial A_{\mu,\nu}}\right)
d{\sigma}_{\mu}=\int_{\sigma}\partial_{\nu }(\sqrt{-g}F^{\mu \nu })d\sigma_{\mu}
=-\int_{S}({\partial}^{\alpha}A^{0})n_{\alpha}r^{2}d\Omega =q\; ,
\end{equation}
where we have used Eq. (\ref{eq:A-asympt}). This implies the \textit{equality} of the total charge of the system and the charge of the source.
\subsection{The case in which $\{{\psi}^{k},b^{k}{}_{\mu},A_{\mu }\}$ is
employed as the set of independent field variables}
\label{subsec:choice2}

Let us denote ${\mbox{\boldmath$L$}}$ and $L$, expressed as functions
of $\psi^{k},b^{k}{}_{\mu},A_{\mu }$ and their derivatives, by 
$\hat{\mbox{\boldmath$L$}}$ and $\hat{L}$, respectively. 
The action ${\mbox{\boldmath$I$}}$ is now written as
\begin{equation}
\label{eq:hat-action}
\hat{\mbox{\boldmath$I$}}=\int_{\cal D}\hat{L}dv= \mbox{\boldmath$I$} \; .
\end{equation}
Various identities can be derived from the requirements 
(R.i) and (R.ii), among which we have
\begin{eqnarray}
\label{eq:identity-hat-1}
\frac{\delta \hat{\mbox{\boldmath$L$}}}{\delta \psi^{k}}&\equiv& 0\;, \\
\label{eq:identity-hat-2}
{}^{\mbox{\rm \footnotesize{tot}}}\hat{\mbox{\boldmath$T$}}_{k}{}^{\mu }&\equiv& 0\; ,
\label{eq:identity-hat-3}
\end{eqnarray}
\begin{eqnarray}
\partial_{\mu }{}^{\mbox{\rm \footnotesize{tot}}}\hat{\mbox{\boldmath$S$}}_{kl}{}^{\mu }
-2\frac{\delta \hat{\mbox{\boldmath$L$}}}{\delta \psi^{[k}}\psi_{l]}
-2\frac{\delta \hat{\mbox{\boldmath$L$}}}{\delta b^{[k}{}_{\mu }}b_{l]\mu}
&\equiv& 0\; ,\\
\label{eq:identity-hat-4}
\hat{\tilde{\mbox{\boldmath$T$}}}_{\mu }{}^{\nu }
-\partial_{\lambda }\hat{\mbox{\boldmath$\Psi $}}_{\mu }{}^{\nu \lambda }
-\frac{\delta \hat{\mbox{\boldmath$L$}}}{\delta b^{k}{}_{\nu }}b^{k}{}_{\mu }
-\frac{\delta \hat{\mbox{\boldmath$L$}}}{\delta A_{\nu }}A_{\mu }&\equiv& 0\; ,
\end{eqnarray}
where we have defined 
\begin{eqnarray}
\label{eq:def-hat-1}
\hat{\mbox{\boldmath$L$}}&\stackrel{\mbox{\scriptsize def}}{=}&
\sqrt{-g}\hat{L}\; ,\\
\label{eq:def-hat-3}
{}^{\mbox{\rm {\footnotesize tot}}}\hat{\mbox{\boldmath$T$}}_{k}{}^{\mu }
&\stackrel{\mbox{\scriptsize def}}{=}&
\frac{\partial \hat{\mbox{\boldmath$L$}}}{\partial \psi^{k}{}_{,\mu }}\; ,\\
\label{eq:def-hat-4}
{}^{\mbox{\rm \footnotesize{tot}}}\hat{\mbox{\boldmath$S$}}_{kl}{}^{\mu }
&\stackrel{\mbox{\scriptsize def}}{=}&
-2\frac{\partial \hat{\mbox{\boldmath$L$}}}{\partial \psi^{[k}{}_{,\mu }}\psi_{l]}
-2\hat{\mbox{\boldmath$F$}}_{[k}{}^{\nu \mu }b_{l]\nu }\; ,\\
\label{eq:def-hat-5}
\hat{\tilde{\mbox{\boldmath$T$}}}_{\mu }{}^{\nu }
&\stackrel{\mbox{\scriptsize def}}{=}&
\delta_{\mu }{}^{\nu }\hat{\mbox{\boldmath$L$}}-\hat{\mbox{\boldmath$F$}}_{k}{}^{\lambda \nu }
b^{k}{}_{\lambda ,\mu }-\hat{\mbox{\boldmath$F$}}^{\lambda \nu }A_{\lambda ,\mu }
-\frac{\partial \hat{\mbox{\boldmath$L$}}}{\partial \psi^{k}{}_{,\nu }}\psi^{k}{}_{,\mu }\;, \\
\label{eq:def-hat-6}
\hat{\mbox{\boldmath$\Psi $}}_{\lambda}{}^{\mu \nu }
&\stackrel{\mbox{\scriptsize def}}{=}&
\hat{\mbox{\boldmath$F$}}_{k}{}^{\mu \nu }b^{k}{}_{\lambda}
+\hat{\mbox{\boldmath$F$}}^{\mu \nu }A_{\lambda }=-\hat{\mbox{\boldmath$\Psi $}}_{\lambda}{}^{\nu \mu }\; ,\\
\label{eq:def-hat-2}
\hat{\mbox{\boldmath$F$}}_{k}{}^{\mu \nu }&\stackrel{\mbox{\scriptsize def}}{=}&
\frac{\partial \hat{\mbox{\boldmath$L$}}}{\partial b^{k}{}_{\mu,\nu }}
={\mbox{\boldmath$F$}}_{k}{}^{\mu \nu }\; ,\; \; \; 
\hat{\mbox{\boldmath$F$}}^{\mu \nu }\stackrel{\mbox{\scriptsize def}}{=}
\frac{\partial \hat{\mbox{\boldmath$L$}}}{\partial A_{\mu,\nu }}
={\mbox{\boldmath$F$}}^{\mu \nu }\; .
\end{eqnarray}
From Eqs. (\ref{eq:identity-hat-1}) and (\ref{eq:identity-hat-3}),
we see that 
\begin{equation}
\label{eq:diff-consv-hat-1}
\partial_{\mu }{}^{\mbox{\rm \footnotesize{tot}}}\hat{\mbox{\boldmath$S$}}_{kl}{}^{\mu }=0\; 
\end{equation}
when the field equations for $b^{k}{}_{\mu }$ are satisfied. From Eqs. (\ref{eq:identity-hat-4})
and (\ref{eq:def-hat-6}), it follows that 
\begin{eqnarray}
\label{eq:diff-consv-hat-2}
\partial_{\nu }\hat{\tilde{\mbox{\boldmath$T$}}}_{\mu }{}^{\nu }&=&0\; ,\\
\label{eq:diff-consv-hat-3}
\partial_{\nu }\hat{\mbox{\boldmath$M$}}_{\lambda }{}^{\mu \nu }&=&0\; 
\end{eqnarray}
when the field equations $\delta \hat{\mbox{\boldmath$L$}}/\delta b^{k}{}_{\mu }=0$ and 
$\delta \hat{\mbox{\boldmath$L$}}/\delta A_{\mu }=0$ are both satisfied, where
$\hat{\mbox{\boldmath$M$}}_{\lambda }{}^{\mu \nu }
\stackrel{\mbox{\scriptsize def}}{=}
2(\hat{\mbox{\boldmath$\Psi $}}_{\lambda }{}^{\mu \nu }-x^{\mu }
\hat{\tilde{\mbox{\boldmath$T$}}}_{\lambda}{}^{\nu })$.
Equations (\ref{eq:diff-consv-hat-1}) 
-- (\ref{eq:diff-consv-hat-3}) are the differential conservation laws of the ``spin'' angular momentum, the canonical angular momentum, and the ``extended orbital angular momentum", respectively. 

The density ${}^{\mbox{\rm \footnotesize{tot}}}\hat{\mbox{\boldmath$S$}}_{kl}{}^{\mu }$ defined by Eq. (\ref{eq:def-hat-4}) can be rewritten as
\begin{eqnarray}
\label{eq:S-hat-alt}
{}^{\mbox{\rm \footnotesize{tot}}}\hat{\mbox{\boldmath$S$}}_{kl}{}^{\mu }&=&
\frac{2}{\kappa }\partial_{\nu }
\left(\sqrt{-g}b^{\mu }{}_{[k}b^{\nu }{}_{l]}
-{}^{(0)}b^{\mu }{}_{[k}{}^{(0)}b^{\nu }{}_{l]}\right)
-b_{[k\nu}{\mbox{\boldmath$F$}}^{(2)}{}_{l]}{}^{\mu \nu }\; ,
\end{eqnarray}
by the use of Eqs. (\ref{eq:F-Z}), (\ref{eq:identity-hat-2}), (\ref{eq:def-hat-2})
and (\ref{eq:def-hat-3}).
\subsubsection{Energy-momentum}
\label{subsubsec:energy-momentum2}
The dynamical energy-momentum $\hat{M}_{k}$, which is the generator of  {\em internal } translations, vanishes identically:
\begin{equation}
\label{eq:internal-m-hat}
\hat{M}_{k}\stackrel{\mbox{\scriptsize def}}{=}
\int_{\sigma}\!^{\mbox{\rm \footnotesize{tot}}}\hat{\mbox{\boldmath$T$}}
_{k}{}^{\mu}d{\sigma}_{\mu}\equiv 0\; .
\end{equation}
This is evident from Eq. (\ref{eq:identity-hat-2}).
\subsubsection{Spin angular momentum}
\label{subsubsec:angular-momentum2}
The generator $\hat{S}_{kl}$ of internal Lorentz
transformations is expressed as
\begin{equation}
\label{eq:s-hat}
\hat{S}_{kl}\stackrel{\mbox{\scriptsize def}}{=}
\int_{\sigma}\!{}^{\mbox{\rm {\footnotesize tot}}} \hat{\mbox{\boldmath$S$}}{}_{kl}{}^{\mu}d{\sigma}_{\mu}
=\frac{2}{\kappa}\int_{S}\left(\sqrt{-g}\,b^{0}{}_{[k}b^{\alpha}
{}_{l]}-{}^{(0)}b^{0}{}_{[k}{}^{(0)}b^{\alpha}{}_{l]}\right)n_{\alpha}
r^{2}d\Omega\; ,
\end{equation}
as can be shown by using Eq. (\ref{eq:S-hat-alt}), and we obtain
\begin{equation}
\label{eq:S-component}
\hat{S}_{(0)a}=0\; , \; \; \; \hat{S}_{ab}=\frac{1}{3}J{\varepsilon}_{ab3}
\end{equation}
by using Eq. (\ref{eq:b-asympt}).
\subsubsection{Canonical energy-momentum and ``extended orbital angular
momentum'' }
\label{subsubsec:canonical2}
The generator $\hat{M}^{c}{}_{\mu}$ of coordinate translations and the
generator $\hat{L}_{\mu}{}^{\nu}$ of $GL(4,\mbox{\boldmath$R$})$ coordinate
transformations are the canonical energy-momentum and the ``extended orbital
angular momentum", respectively. They have the expressions
\begin{eqnarray}
\label{eq:M-L2}
\hat{M}^{c}{}_{\mu}&\stackrel{\mbox{\scriptsize def}}{=}&
\int_{\sigma}
\hat{\tilde{\mbox{\boldmath$T$}}}_{\mu }{}^{\nu }\sigma_{\nu }=
\int_{\sigma}\!\partial_{\tau}\hat{\mbox{\boldmath$\Psi$}}_{\mu}{}^{\nu\tau}
d{\sigma}_{\nu}\; ,\\
\label{eq:acm}
\hat{L}_{\mu}{}^{\nu}&\stackrel{\mbox{\scriptsize def}}{=}&
\int_{\sigma}\hat{\mbox{\boldmath$M$}}_{\mu }{}^{\nu \lambda }d{\sigma}_{\lambda}=
-2\int_{\sigma}\!\partial_{\tau}\left(x^{\nu}
\hat{\mbox{\boldmath$\Psi $}}_{\mu}{}^{\lambda\tau}\right)d{\sigma}_{\lambda}\; .
\end{eqnarray}
Then we have  
\begin{equation}
\label{eq:hat-canonical}
\hat{M}^{c}{}_{0}=-m\; , \; \; \; \hat{M}^{c}{}_{\alpha}=0\; . 
\end{equation}
Thus, $\hat{M}^{c}{}_{\mu}$ is the total energy-momentum, and the
\textit{equality} of the active gravitational mass and the inertial mass
holds. 
Also, $\hat{L}_{\mu}{}^{\nu}$ is given by
\begin{eqnarray}
\label{eq:hat-L}
\hat{L}_{0}{}^{0}&=&2x^{0}m\;, \; \; \; \hat{L}_{0}{}^{\alpha}=0\; ,\; \; \; 
\hat{L}_{\alpha}{}^{0}=0\; ,\nonumber \\
\hat{L}_{\alpha}{}^{\beta}&=&\frac{2}{3}J{\varepsilon}_{\alpha}{}^{\beta3}\; , \; \; (\alpha \neq \beta )\; ,\; \; \; \hat{L}_{1}{}^{1}=\hat{L}_{2}{}^{2}=\hat{L}_{3}{}^{3}=\infty \; .
\end{eqnarray}
Equations (\ref{eq:hat-canonical}) and (\ref{eq:hat-L}) are obtained by using
Eqs. (\ref{eq:b-asympt}), (\ref{eq:A-asympt}), (\ref{eq:F02}), (\ref{eq:def-hat-6}), (\ref{eq:def-hat-2}) and (\ref{eq:asympt-F}). 

The orbital angular momentum $\hat{L}_{[\mu \nu ]}$ is given by
\begin{equation}
\label{eq:anti-L-hat}
\hat{L}_{[0 \alpha ]}=0\; ,\; \; \; 
\hat{L}_{[\alpha \beta]}=\frac{2}{3}J{\varepsilon}_{\alpha \beta3}
\; .
\end{equation}
If we define the total angular momentum of the system by\cite{Kawai-Saitoh}
\begin{equation}
\label{eq:J}
\hat{J}_{kl}\stackrel{\mbox{\scriptsize def}}{=}
\hat{S}_{kl}+{}^{(0)}b^{\mu}{}_{k}{}^{(0)}b^{\nu}{}_{l}
\hat{L}{}_{[\mu \nu]}\; ,
\end{equation}
then we have 
\begin{equation}
\label{eq:J-component}
\hat{J}_{(0)k}=0\; ,\; \; \; \hat{J}_{ab}=J{\varepsilon}_{ab3}\; ,
\end{equation}
and the total angular momentum is equal to the angular momentum of the 
rotating source.
\subsubsection{Charge}
\label{subsubsec:charge2}

The generator $\hat{C}$ of $U(1)$ gauge transformations is given by 
\begin{equation}
\label{eq:c-hat}
\hat{C}\stackrel{\mbox{\scriptsize def}}{=}\int_{\sigma}\!{\partial}_{\nu}\left(\frac{\partial\hat{\mbox{\boldmath$L$}}}{\partial A_{\mu,\nu}}\right)
d{\sigma}_{\mu }=q\; ,
\end{equation}
which implies the \textit{equality} of the total charge of the system and
the charge of the source.

\section{Restrictions imposed on field variables \\ by the generalized equivalence principle}
\label{sec:restriction}
In the preceding section, we examined the solution given by Eqs.~(\ref{eq:v-fields}) and (\ref{eq:em-potential}), and the 
results show that the total energy-momentum, the total angular momentum, 
and the total charge of the system, which are generators of 
transformations, are equal to the corresponding active quantities 
of a central gravitating body. 
The total mass, which is equal to the total energy divided by 
the square of the velocity of light, can be regarded as the inertial 
mass of the system. Thus, the results include the equality 
of the inertial mass and the active gravitational mass, which implies 
that the equivalence principle is satisfied by this solution.

In view of the above, we regard, the total momentum, the total 
angular momentum and the total charge as \lq \lq inertial" quantities, 
and we say that a generalized equivalence principle (G. E. P.) is satisfied 
if the total energy-momentum, total angular momentum and total charge are 
all equal to the corresponding quantities of the source. 

The axial vector part $a_{\mu}$ vanishes for our solution, as stated above,
and the field equations (\ref{eq:G-eq}) -- (\ref{eq:EM-eq}) are covariant under general coordinate transformations and under local Lorentz transformations that keep $a_{\mu}$ vanishing. 
Thus, new solutions can be obtained by applying the general coordinate transformations and restricted local Lorentz transformations $b^{k\prime }{}_{\mu}=A^{k}{}_{l}(x)b^{l}{}_{\mu }$ that satisfy the condition
\cite{Hayashi-Shirafuji}${}^{,}$\footnote{Note that $a_{\mu}$ is invariant under the local Lorentz transformation $A^{k}{}_{l}(x)$ if and only if this condition is satisfied.}
\begin{equation}
\label{eq:restricted-local-Lorentz}
\varepsilon^{\mu \nu \rho \sigma }
b^{k}{}_{\nu }b^{l}{}_{\rho }A^{m}{}_{k}(x)A_{ml}(x){}_{,\sigma }=0
\end{equation}
to the solution represented by Eqs. (\ref{eq:v-fields}) and (\ref{eq:em-potential}).

In this section, we examine restrictions imposed on solutions 
by the requirement that this G. E. P. is satisfied. We look for new solutions having suitable asymptotic behavior by considering  the following $\overline{\mbox{Poincar\'e}}$ gauge 
transformations:\\
(1) Local\thinspace$SL(2.C)$ transformation
\begin{equation}
\label{eq:sl2ctr}
H^{k}{}_{l}\stackrel{\mbox{\scriptsize def}}{=}
{\left(\Lambda(a^{-1})\right)}^{k}{}_{l}
={\Lambda}^{k}{}_{l}+{\Lambda}
^{k}{}_{m}{\omega}^{m}{}_{l}(x)\; .
\end{equation}
(2) Local internal translation
\begin{equation}
\label{eq:inttr}
t^{k}={}^{(0)}t^{k}+b^{k}(x)\; .
\end{equation}
Here, ${\Lambda}^{k}{}_{l}$ and ${}^{(0)}t^{k}$ denote a constant internal Lorentz transformation and 
the constant internal translation, respectively, and
${\omega}^{k}{}_{l}$ and $b^{k}$ are functions satisfying the following conditions:
\begin{eqnarray}
\label{eq:omega1}
{\omega}^{k}{}_{l}(x)&=&O^{k}{}_{l}\left(\displaystyle\frac{\mathstrut1}
{r^{p}}\right)\; , \; \; \;  {\omega}^{k}{}_{l,\mu}(x)=O^{k}{}_{l\mu}\left(\displaystyle\frac{\mathstrut1}
{r^{p+1}}\right)\; ,\>(p>0)\; \\
\label{eq:b}
b^{k}(x)&=&O^{k}{}_{{}}\left(\displaystyle\frac{\mathstrut1}{r^{\gamma}
}\right)\; , \; \; \; b^{k}{}_{,\mu }(x)=O^{k}{}_{\mu }
\left(\displaystyle\frac{\mathstrut1}{r^{\gamma+1}}\right)\; . \>(\gamma>0)\; 
\end{eqnarray}
The transformation $H^{k}{}_{l}\stackrel{\mbox{\scriptsize def}}{=}
{\left(\Lambda(a^{-1})\right)}^{k}{}_{l}$ is a Lorentz transformation satisfying \linebreak
Eq.~(\ref{eq:restricted-local-Lorentz}), if and only if
\begin{eqnarray}
\label{eq:restriction-omega-1}
& &\omega_{kl}+\omega_{lk}+\omega_{mk}\omega^{m}{}_{l}=0\; ,\\
\label{eq:restriction-omega-2}
& &\varepsilon^{\mu \nu \lambda \rho}(\omega_{kl,\rho}
+\omega_{mk}\omega^{m}{}_{l,\rho})b^{k}{}_{\nu }b^{l}{}_{\lambda }=0\; 
\end{eqnarray}
are both satisfied. 
The conditions (\ref{eq:restriction-omega-1}) and (\ref{eq:restriction-omega-2}) are equivalent to 
\begin{equation}
\label{eq:omega-1G}
\omega_{\mu \nu }+\omega_{\nu \mu }
+\omega_{\lambda \mu }\omega^{\lambda }{}_{\nu }=0
\end{equation}
and 
\begin{equation}
\label{eq:omega-2G}
X_{\mu \nu \lambda}+X_{\lambda \mu \nu }+X_{\nu \lambda \mu }
=0\; ,
\end{equation}
respectively, where we have defined
\begin{eqnarray}
\label{eq:omega-G}
\omega_{\mu \nu }
&\stackrel{\mbox{\scriptsize def}}{=}&
b^{k}{}_{\mu }b^{l}{}_{\nu }\omega_{kl}\; ,\\
\label{eq:X}
X_{\mu \nu \lambda }&\stackrel{\mbox{\scriptsize def}}{=}&
\omega_{\mu \nu ,\lambda }
+\omega^{\tau}{}_{\mu }\omega_{\tau \nu ,\lambda}
-b^{\tau }{}_{k}b^{k}{}_{\mu,\lambda}
\omega_{\tau \nu}-b^{\tau}{}_{k}b^{k}{}_{\nu ,\lambda}\omega_{\mu \tau }
\nonumber \\
& &+b^{\tau k}b^{\rho }{}_{k,\lambda }\omega_{\tau \mu }
\omega_{\rho \nu}+b^{k}{}_{\nu }b^{\tau }{}_{k,\lambda}
\omega^{\rho}{}_{\mu }\omega_{\rho \tau }\; .
\end{eqnarray}
The function $X_{\mu \nu \lambda }$ is anti-symmetric with respect to the first two indices:
\begin{equation}
\label{eq:X-antisymmetry}
X_{\mu \nu \lambda}=-X_{\nu \mu \lambda }\; . 
\end{equation}
From Eqs. (\ref{eq:omega1}), (\ref{eq:omega-1G}) and (\ref{eq:omega-2G}), 
${\omega}_{\mu \nu }$ is known to have the expression
\begin{eqnarray}
\label{eq:omega2}
\omega_{\mu\nu}&=&
\partial_{\mu}\omega_{\nu}-\partial_{\nu}\omega_{\mu}+f_{\mu\nu}(x)\; ,
\end{eqnarray}
with
\begin{eqnarray}
\label{eq:omega3}
\partial_{\mu}\omega_{\nu}-\partial_{\nu}\omega_{\mu}&=&
O_{[\mu\nu]}\left(\displaystyle\frac{\mathstrut1}{r^{p}}\right)\; ,\nonumber \\
f_{\mu\nu}(x)&=&O_{\mu\nu}\left(\displaystyle\frac{\mathstrut1}
{r^{s}}\right) \; . \; (\,p<s\,)\; 
\end{eqnarray}
In addition, we require the leading term of $\omega_{kl}$ at spatial infinity
to be spherically symmetric. Then we can write 
\begin{equation}
\label{eq:spherical}
\omega^{0}=A(r,x^{0})\; ,\quad\omega^{\alpha}=n^{\alpha}B(r,x^{0})\; , 
\end{equation}
with certain functions $A$ and $B$ of $r$ and $x^{0}$.

Also, we consider the following coordinate transformation:
\begin{eqnarray}
\label{eq:coordinate}
x^{\prime\mu}&=&C^{\mu}{}_{\nu}x^{\nu}+D^{\mu}(x)\;,\nonumber \\
\displaystyle\frac{\mathstrut\partial x^{\prime\mu}}{\partial x^{\nu}
}&=&C^{\mu}{}_{\nu}+a^{\mu}{}_{\nu}(x)\; ,\nonumber \\
a^{\mu}{}_{\nu}&\stackrel{\mbox{\scriptsize def}}{=}&D^{\mu}{}_{,\nu}\; ,
\nonumber \\
a^{\mu}{}_{\nu}(x)&=&O^{\mu}{}_{\nu}\left(\displaystyle\frac{\mathstrut
1}{r^{u}}\right)  ,\quad a^{\mu}{}_{\nu,\lambda}=
O^{\mu}{}_{\nu \lambda}\left(\displaystyle\frac{\mathstrut1}{r^{u+1}}\right)  ,\quad (u>0)\; 
\end{eqnarray}
where $C^{\mu}{}_{\nu}$ denotes a constant Lorentz transformation, and
$\mathstrut D^{\mu}(x)$\ satisfies the condition
\begin{equation}
\label{eq:lim-A}
\lim_{r\rightarrow\infty}\displaystyle\frac{\mathstrut D^{\mu}(x)}
{r}=0\; .
\end{equation}
We write $\partial x^{\mu }/\partial x^{\prime\nu}$ as
\begin{equation}
\label{eq:inverse-coordinate}
\displaystyle\frac{\mathstrut\partial x^{\mu}}{\partial x^{\prime\nu}
}=\left(C^{-1}\right)^{\mu}{}_{\nu}+d^{\mu}{}_{\nu}(x)\; ,
\end{equation}
with $(C^{-1})^{\mu }{}_{\nu }$ being constants satisfying $(C^{-1})^{\mu }{}_{\lambda}C^{\lambda }{}_{\nu}=\delta^{\mu }{}_{\nu }$.

The vierbeins and vector potentials given by 
\begin{equation}
\label{eq:new-solution}
b^{\prime k }{}_{\mu}\stackrel{\mbox{\scriptsize def}}{=}
H^{k}{}_{l}\frac{\partial x^{\nu }}
{\partial x^{\prime\mu }}b^{l}{}_{\nu }\; ,\; \; \; 
A^{\prime}{}_{\mu }\stackrel{\mbox{\scriptsize def}}{=}
\frac{\partial x^{\nu }}{\partial x^{\prime\mu }}A_{\nu }\; ,
\end{equation}
with $b^{k}{}_{\mu }$ and $A_{\mu }$ given by Eqs. (\ref{eq:v-fields}) and 
(\ref{eq:em-potential}), are solutions of the gravitational and electromagnetic 
field equations. This is true irrespective of the values of the parameters 
$p\; ,s\; ,\beta\; ,\gamma $ and $u$. The G. E. P. is considered to be satisfied if the energy-momentum, the angular momentum and the charge all have correct transformation properties as their indices indicate. 
But, this is not necessarily the case for arbitrary values of these parameters; i.e. 
there are solutions which do not satisfy the G. E. P. We examine restrictions imposed on these parameters by the requirement 
that the G. E. P. is satisfied.
\subsection{The case in which $\{{\psi}^{k},A^{k}{}_{\mu},A_{\mu }\}$ is employed as the set of independent field variables}
\label{subsec:choice12}
Under the combined transformation of the $\overline{\mbox{Poincar\'e}}$ gauge transformation given by Eqs.~(\ref{eq:sl2ctr}) and (\ref{eq:inttr}) and satisfying the conditions
(\ref{eq:restriction-omega-1}) and (\ref{eq:restriction-omega-2}) 
and of the coordinate transformation (\ref{eq:coordinate}), 
${\mbox{\boldmath$F$}}^{(1)}{}_{k}{}^{\mu\nu}, 
{\mbox{\boldmath$F$}}^{(2)}{}_{k}{}^{\mu\nu}$ and ${\mbox{\boldmath$F$}}^{\mu \nu }$
transform according to 
\begin{eqnarray}
\label{eq:F1-prime}
{\mbox{\boldmath$F$}}^{(1)\prime}{}_{k}{}^{\mu \nu}&=&
\Delta\displaystyle\frac{\mathstrut \partial x^{\prime\mu}}
{\partial x^{\rho}}
\displaystyle\frac{\mathstrut \partial x^{\prime\nu}}
{\partial x^{\sigma}}{}H_{k}{}^{l}
{\mbox{\boldmath$F$}}^{(1)}{}_{l}{}^{\rho\sigma}\nonumber \\
& &+\;\frac{\Delta}{\kappa}U_{k}{}^{lmn}{}_{\lambda}
W^{\rho \sigma \lambda}{}_{lmn}
\displaystyle \frac{\mathstrut\partial x^{\prime\mu}}
{\partial x^{\rho}}
\displaystyle \frac{\mathstrut\partial x^{\prime\nu}}
{\partial x^{\sigma}}\; ,\\
\label{eq:F2-prime}
{\mbox{\boldmath$F$}}^{(2)\prime}{}_{k}{}^{\mu \nu}
&=&0=\Delta\displaystyle\frac{\mathstrut \partial x^{\prime\mu}}
{\partial x^{\rho}}
\displaystyle\frac{\mathstrut \partial x^{\prime\nu}}
{\partial x^{\sigma}}{}H_{k}{}^{l}
{\mbox{\boldmath$F$}}^{(2)}{}_{l}{}^{\rho\sigma}\; ,\\
\label{eq:emF-prime}
{\mbox{\boldmath$F$}}^{\prime \mu \nu}
&=&\Delta \frac{\partial x^{\prime\mu}}{\partial x^{\rho }}
\frac{\partial x^{\prime\nu }}{\partial x^{\sigma }}{\mbox{\boldmath$F$}}^{\rho \sigma }\; ,
\end{eqnarray}
where we have defined\footnote{For simplicity, we restrict our consideration to the case in which $\Delta>0$. An extension to the case of arbitrary non-vanishing $\Delta $ can be made without difficulty.}   
\begin{eqnarray}
\label{eq:U-V-W-delta}
U_{k}{}^{lmn}{}_{\lambda}&\stackrel{\mbox{\scriptsize def}}{=}&
H_{k}{}^{l}V^{mn}{}_{\lambda}\; ,\\
V^{mn}{}_{\lambda }&\stackrel{\mbox{\scriptsize def}}{=}&
H^{lm}H_{l}{}^{n}{}_{,\lambda}=-V^{nm}{}_{\lambda }\; ,\\
W^{\mu\nu\lambda}{}_{klm}
&\stackrel{\mbox{\scriptsize def}}{=}&
b^{[\mu}{}_{k}b^{\nu]}{}_{l}b^{\lambda}{}_{m}
+b^{[\mu}{}_{m}b^{\nu]}{}_{k}b^{\lambda}
{}_{l}+b^{[\mu}{}_{l}b^{\nu]}{}_{m}b^{\lambda}{}_{k}\,,\\
\Delta&\stackrel{\mbox{\scriptsize  def}}{=}&
\det\left(\displaystyle\frac{\mathstrut
\partial x^{\mu}}{\partial x^{\prime \nu }}\right) \; .
\end{eqnarray}
Equations (\ref{eq:F1-prime})
~-- (\ref{eq:emF-prime}) 
show that ${\mbox{\boldmath$F$}}^{(1)}{}_{k}{}^{\mu\nu}\;,{\mbox{\boldmath$F$}}^{(2)}{}_{k}{}^{\mu \nu}$ and ${\mbox{\boldmath$F$}}^{\mu \nu }$ transform as tensor 
densities under coordinate transformations. The function
$W^{\mu\nu\lambda}{}_{klm}$ is totally anti-symmetric, both in upper 
indices and in lower indices.
\subsubsection{Energy-momentum}
\label{subsubsec:energy-momentum12}
From Eqs.~(\ref{eq:energy-momentum-alt}) and (\ref{eq:F1-prime}), $M_{k}$ is found to transform as\footnote{$A{}^{[\mu\nu\lambda]}\stackrel{\mbox{\scriptsize def}}{=}
\frac{1}{3}\{A{}^{\mu\lbrack\nu\lambda]}+A{}^{\nu\lbrack\lambda\mu]}+A{}^{\lambda\lbrack\mu\nu]}\}$.}
\begin{equation}
\label{eq:energy-mementum-trans1}
M^{\prime}{}_{k}\stackrel{\mbox{\scriptsize def}}{=}
\int_{\sigma}{}^{\mbox{\rm \footnotesize{tot}}}{\mbox{\boldmath$T$}}^{\prime}{}_{k}{}^{\mu}
d\sigma^{\prime}{\hspace{0.01mm}}_{\mu}=
\int_{\sigma }\partial^{\prime}{}_{\nu}
{\mbox{\boldmath$F$}}^{\prime}{}_{k}{}^{\mu \nu}d\sigma^{\prime}{\hspace{0.01mm}}_{\mu}
=\Lambda_{k}{}^{l}M_{l}
+\frac{3}{\kappa}\int U_{k}{}^{[0\alpha\beta]}{}_{\beta}n_{\alpha}r^{2}
d\Omega \; ,
\end{equation}
where we represent $\partial^{\prime}{\hspace{0.01mm}}_{\mu} = \partial/\partial x^{\prime\mu}.$
The energy-momentum $M_{k}$ obeys the correct transformation rule
\begin{equation}
\label{eq:energy-momentum-trans2}
M^{\prime}{}_{k}=\Lambda_{k}{}^{l}M_{l}\; 
\end{equation}
if the condition
\begin{equation}
\label{eq:mk-transcond}
p>\frac{1}{2}\; ,\; \; \; s>1 
\end{equation}
is satisfied.\footnote{For derivations of the conditions 
(\ref{eq:mk-transcond}), (\ref{eq:s-condi-01}) 
-- (\ref{eq:s-condi-03}), (\ref{eq:L-cond}), (\ref{eq:m-hat-cond}) and (\ref{eq:L-J-hat-cond}),  elementary but rather tedious calculations are needed. In Appendix A, we give lists of asymptotic forms of ${\mbox{\boldmath$F$}}^{(1)}{}_{k}{}^{\mu \nu },
V^{mn}{}_{\lambda}$ and $W^{\mu \nu \lambda }{}_{klm}$ for large $r$, which are useful in calculations.}
\subsubsection{Angular momentum}
\label{subsubsec:angular-momentum12}
From Eqs. (\ref{eq:total-angular-momentum}) and (\ref{eq:F1-prime}), we find that $S_{kl}$ transforms as
\begin{eqnarray}
\label{eq:S-transform}
S^{\prime}{}_{kl}
&\stackrel{\mbox{\scriptsize def}}{=}&
\int_{\sigma}{}^{\mbox{\rm \footnotesize{tot}}}{\mbox{\boldmath$S$}}^{\prime}{}_{kl}{}^{\mu}
d\sigma^{\prime}{\hspace{0.01mm}}_{\mu}\nonumber \\
&=&2\int_{\sigma }\partial^{\prime}{\hspace{0.01mm}}_{\nu}
\left[\psi^{\prime}{}_{[k}{\mbox{\boldmath$F$}}^{\prime}{}_{l]}{}^{\mu \nu}
+\frac{1}{\kappa }
\left(\sqrt{-g^{\prime }}b^{\prime\mu }{}_{[k}
b^{\prime\nu }{}_{l]}
-b^{(0)\prime\mu }{}_{[k}b^{(0)\prime\nu }{}_{l ]}
\right)\right]d\sigma^{\prime}{\hspace{0.01mm}}_{\mu}\nonumber \\
&=&\Lambda_{k}{}^{m}\Lambda_{l}{}^{n}(S_{mn}-2{}^{(0)}t_{[m}
M_{n]})\nonumber \\
& &+2\Lambda_{k}{}^{m}\int \{\omega_{m}{}^{n}(\psi_{n}-t_{n})
-b_{m}\}
H_{l}{}^{n}{\mbox{\boldmath$F$}}_{n}{}^{0\alpha }n_{\alpha }r^{2}d\Omega \nonumber \\
& &+\frac{2}{\kappa }\int H_{[k}{}^{m}H_{l]}{}^{n}(\psi_{m}-t_{m})
H^{ij}H_{i}{}^{h}{}_{,\beta}W^{0\alpha \beta }{}_{njh}n_{\alpha }r^{2}d\Omega \nonumber \\
& &+\frac{2}{\kappa }\int H_{[k}{}^{m}(H_{l]}{}^{n}
+\Lambda_{l]}{}^{m}\omega_{m}{}^{n})(b^{0}{}_{[m}b^{\alpha }{}_{n]}
-{}^{(0)}b^{0}{}_{[m}{}^{(0)}b^{\alpha }{}_{n]})n_{\alpha }r^{2}d\Omega \; ,\nonumber \\
{}
\end{eqnarray}
where we have defined ${}^{(0)}b^{\prime\mu}{}_{k}
\stackrel{\mbox{\scriptsize  def}}{=}\lim_{r\rightarrow \infty}
H_{k}{}^{l}(\partial x^{\prime\mu }/\partial x^{\nu })
b^{\nu }{}_{l}=
\Lambda_{k}{}^{l}C^{\mu }{}_{\nu }{}^{(0)}b^{\nu }{}_{l}$. 
The angular momentum obeys the correct transformation rule\footnote{The term $2{}^{(0)}t_{[k}M_{l]}\;$ comes from the translation (\ref{eq:inttr}), and it is an additional orbital angular momentum around the origin of the
\emph{internal\/} space.}
\begin{equation}
\label{eq:s-trans}
S^{\prime}{}_{kl}=\Lambda_{k}{}^{m}\Lambda_{l}{}^{n}
(S_{mn}-2{}^{(0)}t_{[m}M_{n]})\; 
\end{equation}
if the following conditions are satisfied: 
\begin{equation}
\label{eq:s-condi-01}
p>\frac{1}{2}\; ,\; \; \; s>2\; 
\end{equation}
and
\begin{equation}
\label{eq:s-condi-02}
\left[\{\;p+\beta>1,\quad p+\gamma>1 \}\; \; {\mbox{\rm or}}\; \; 
\left\{\;O^{(0)}\left(\displaystyle\frac{1}{r^{\beta}}\right)=
f(r,x^{0})\;,\;O^{(0)}
{}\left(\displaystyle\frac{1}{r^{\gamma}}\right)=
g(r,x^{0})\;\right \}\right]
\end{equation}
and
\begin{equation}
\label{eq:s-condi-03}
\left[\{\;p+\beta>1,\; p+\gamma>1 \}\;  {\mbox{\rm or}}\; \nonumber 
\left\{\;O^{a}\left(\displaystyle\frac{1}{r^{\beta}}\right)\,=
n^{a}h(r,x^{0})\; ,\; O^{a}\left(\displaystyle\frac{1}{r^{\gamma}}\right)
=n^{a}k(r,x^{0})\;\right\}\right]\;,
\end{equation}
with $f,g,h$ and $k$ being some functions of $r$ and $x^{0}$, where the terms $O^{k}(1/r^{\beta })\;$ and $O^{k}(1/r^{\gamma})\;$ are those in 
Eq. (\ref{eq:psihyou}) and Eq. (\ref{eq:b}), respectively.
\subsubsection{Canonical energy-momentum and ``extended orbital 
angular momentum''}
\label{subsubsec:canonical12}
The transformed canonical energy-momentum $M^{\prime c}{}_{\mu}$ and the
``extended orbital angular momentum" $L^{\prime}{}_{\mu}{}^{\nu}$
are given by 
\begin{eqnarray}
\label{eq:trans-CEM-EAM}
M^{\prime c}{}_{\mu}
&\stackrel{\mbox{\scriptsize def}}{=}&
\int_{\sigma }\tilde{\mbox{\boldmath$T$}}^{\prime}{}_{\mu}
{}^{\nu}d\sigma^{\prime}{\hspace{0.01mm}}_{\nu}
=\int_{\sigma}\partial^{\prime}{\hspace{0.01mm}}_{\lambda}
{\mbox{\boldmath$\Psi $}}^{\prime}{}_{\mu}{}^{\nu \lambda}
d\sigma^{\prime}{\hspace{0.01mm}}_{\nu}\nonumber \\
&=& \int {\mbox{\boldmath$\Psi$}}^{\prime}{}_{\mu}{}^{\nu \lambda}
J\frac{\partial x^{0}}{\partial x^{\prime\nu}}
\frac{\partial x^{\alpha }}{\partial x^{\prime\lambda}}n_{\alpha }r^{2}
d\Omega \; , \\
L^{\prime}{}_{\mu}{}^{\nu}
&\stackrel{\mbox{\scriptsize def}}{=}&
\int_{\sigma}{\mbox{\boldmath$M$}}^{\prime}{}_{\mu}{}^{\nu\lambda}
d{\sigma }^{\prime}{\hspace{0.01mm}}_{\lambda}
=-2\int_{\sigma}\partial^{\prime}{\hspace{0.01mm}}_{\tau}(x^{\prime\nu }{\mbox{\boldmath$\Psi $}}^{\prime}_{\mu }{}^{\lambda \tau})d\sigma^{\prime}{\hspace{0.01mm}}_{\lambda}\nonumber \\
&=&-2\int x^{\prime\nu }{\mbox{\boldmath$\Psi$}}^{\prime}{}_{\mu}{}^{\lambda\tau}J
\frac{\partial x^{0}}{\partial x^{\prime\lambda}}
\frac{\partial x^{\alpha }}{\partial x^{\prime\tau}}n_{\alpha }r^{2}
d\Omega \; ,
\end{eqnarray} 
with $J\stackrel{\mbox{\scriptsize def}}{=}
\det(\partial x^{\prime\mu }/\partial x^{\nu })$, in which the transformed  
${\mbox{\boldmath$\Psi $}}^{\prime}{}_{\mu}{}^{\nu \lambda}$ can be obtained
from Eqs. (\ref{eq:transformation}), (\ref{eq:def-6}), (\ref{eq:F1-prime}) 
-- (\ref{eq:emF-prime}).

The relation 
\begin{equation}
\label{eq:CEM-trans}
M^{\prime c}{}_{\mu}=0=(C^{-1})^{\nu }{}_{\mu}M^{c}{}_{\nu }
\end{equation}
holds without any additional condition on the positive parameters 
$p\; ,s\; ,\beta \;, \gamma $ and $u$, and 
$L_{\mu }{}^{\nu }$ and $L_{[\mu \nu]}$ transform according to
\begin{eqnarray}
\label{eq:L-trans}
L^{\prime}{}_{\mu }{}^{\nu }&=&(C^{-1})^{\rho}{}_{\mu }C^{\nu}{}_{\sigma }L_{\rho }{}^{\sigma}\; ,\\
\label{eq:L-orbit-trans}
L^{\prime}{}_{[\mu \nu ]}&=&0=(C^{-1})^{\lambda }{}_{\mu }(C^{-1})^{\rho}{}_{\nu }L_{[\lambda \rho]}\; 
\end{eqnarray}
under the condition 
\begin{equation}
\label{eq:L-cond}
p>1\; .
\end{equation}
Equations (\ref{eq:CEM-trans})
-- (\ref{eq:L-orbit-trans}) are the transformation rules which we would like $M^{c}{}_{\mu }, L_{\mu }{}^{\nu}$ and 
$L_{[\mu \nu]}$ to obey. 
\subsubsection{Charge}
\label{subsubsec:charge12}
For the transformed charge
\begin{equation}
\label{eq:charge-trans} 
C^{\prime }\stackrel{\mbox{\scriptsize def}}{=}
\int_{\sigma}\!\partial^{\prime}{\hspace{0.01mm}}_{\nu}
\left(\frac{\partial \mbox{\boldmath$L^{\prime }$}}
{\partial A^{\prime}{}_{\mu,\nu}}\right)d{\sigma}^{\prime}{\hspace{0.01mm}}_{\mu}\; ,
\end{equation}
with the Lagrangian density $\mbox{\boldmath$L^{\prime }$}$ defined with 
transformed field variables, we can obtain   
\begin{equation}
\label{eq:charge-trans-eq}
C^{\prime}=C=q\; 
\end{equation}
without imposing any additional condition.

To summarize, the G. E. P. is satisfied if 
\begin{equation}
\label{eq:parameter-condition}
p>1\;,\; \; \; s>2\; .
\end{equation}
The asymptotic behavior of the transformed dual components of 
$b^{\prime k}{}_{\mu}= \linebreak 
H^{k}{}_{l}(\partial x^{\nu }/\partial x^{\prime \mu })
b^{l}{}_{\nu }$ and of the transformed vector potential 
$A^{\prime}{}_{\mu}=(\partial x^{\nu }/\partial x^{\prime\mu })A_{\nu }$ can be
easily obtained from Eqs. (\ref{eq:sl2ctr}), (\ref{eq:omega1}), 
(\ref{eq:inverse-coordinate}) and (\ref{eq:parameter-condition})
as 
\begin{eqnarray}
\label{eq:b-A-prime-asympt}
b^{\prime k }{}_{\mu }
&=&\Lambda^{k}{}_{l}(C^{-1})^{\nu }{}_{\mu }{}^{(0)}b^{l}{}_{\nu }+
O^{k}{}_{\mu }(1/r^{u})+O^{k}{}_{\mu }(1/r)\; , \; \; \; (u>0)\; \nonumber \\
A^{\prime}{}_{\mu }&=&
(C^{-1})^{\nu }{}_{\mu }A_{\nu }+d^{\nu }{}_{\mu }A_{\nu }=O_{\mu }(1/r)\; .
\end{eqnarray}
\subsection{The case in which $\{{\psi}^{k},b^{k}{}_{\mu},A_{\mu }\}$ is employed as the set of independent field variables}
\label{subsec:choice22}

\subsubsection{Energy-momentum and angular momentum}
\label{subsubsec:energy-angular-momentum}

For the generator of $\overline{\mbox{Poincar\'e}}$ gauge
transformations, we have 
\begin{eqnarray}
\label{eq:hat-em-trans}
\hat{M}^{\prime}{}_{k}&\stackrel{\mbox{\scriptsize def}}{=}&
\int_{\sigma}{}^{\mbox{\rm \footnotesize{tot}}}\hat{{\mbox{\boldmath$T$}}^{\prime}}{}_{k}{}^{\mu}
d\sigma^{\prime}{\hspace{0.01mm}}_{\mu}
\equiv 0=\Lambda_{k}{}^{l}\hat{M}_{l}\; , \\ 
\label{eq:hat-am-trans}
\hat{S}^{\prime}{}_{k l}&\stackrel{\mbox{\scriptsize def}}{=}&
\int_{\sigma}{}^{\mbox{\rm \footnotesize{tot}}}\hat{{\mbox{\boldmath$S$}}^{\prime}}{}_{k l}{}^{\mu}
d\sigma^{\prime}{\hspace{0.01mm}}_{\mu}={\Lambda}_{k}{}^{m}{\Lambda}_{l}{}^{n}\hat{S}_{mn}\; ,
\end{eqnarray}
which hold without imposing any additional condition.
\subsubsection{Canonical energy-momentum and ``extended orbital angular momentum"}
\label{subsubsec:canonical22}
For $\hat{M^{c}}{}_{\mu}$, we have 
\begin{equation}
\label{eq:m-hat}
\hat{M}^{\prime c}{}_{\mu}
\stackrel{\mbox{\scriptsize def}}{=}
\int_{\sigma}\hat{\tilde{\mbox{\boldmath$T$}^{\prime}{}}}_{\mu}{}^{\nu}
d\sigma^{\prime}{\hspace{0.01mm}}_{\nu}=(C^{-1})^{\nu}{}_{\mu}\hat{M}^{c}{}_{\nu}\; 
\end{equation}
if the conditions
\begin{equation}
\label{eq:m-hat-cond}
p>\frac{1}{2}\; ,\; \; \; s>1\; ,\; \; \; p+u>1
\end{equation}
are satisfied. The transformed \lq \lq extended orbital angular momentum" 
\begin{equation}
\label{eq:L-hat}
\hat{L}^{\prime}{}_{\mu }{}^{\nu}
\stackrel{\mbox{\scriptsize def}}{=}\int_{\sigma }\hat{\mbox{\boldmath$M$}^{\prime}}{}_{\mu}
{}^{\nu \lambda}d{\sigma }^{\prime}{\hspace{0.01mm}}_{\lambda} 
\end{equation}
is divergent in general, as is obvious from
Eq. (\ref{eq:hat-L}). However, the transformed 
orbital angular momentum $\hat{L}^{\prime}{}_{[\mu \nu]}$, and 
hence the transformed total angular momentum 
$\hat{J}^{\prime}{}_{k l}
\stackrel{\mbox{\scriptsize def}}{=}\hat{S}^{\prime}{}_{k l}
+{}^{(0)}b^{\prime \mu}{}_{k}{}^{(0)}b^{\prime \nu}{}_{l}
\hat{L}^{\prime}{}_{[\mu\nu]}$, are well defined, and they 
obey the rules
\begin{eqnarray}
\label{eq:orbital-L-hat-trans}
\hat{L}^{\prime}{}_{[\mu\nu]}
&=&(C^{-1})^{\lambda}{}_{\mu}(C^{-1})^{\rho}{}_{\nu}
\hat{L}_{[\lambda \rho]}\; ,\\
\label{eq:anti-J-trans}
\hat{J}^{\prime}{}_{kl}
&=&{\Lambda}_{k}{}^{m}{\Lambda}_{l}{}^{n}\hat{J}_{mn}\; 
\end{eqnarray}
if the conditions 
\begin{equation}
\label{eq:L-J-hat-cond}
p>\frac{1}{2}\; ,\; \; \; s>2\; ,\; \; \; u>1\; ,\; \; \; p+u>2 
\end{equation}
are satisfied. 
\subsubsection{Charge}
\label{subsub:charge22}
The transformed charge
\begin{equation}
\label{eq:hat-charge-trans} 
\hat{C}^{\prime }
\stackrel{\mbox{\scriptsize def}}{=}\int_{\sigma}\!{\partial}{\hspace{0.01mm}}^{\prime}{}_{\nu}
\left(\frac{\partial \hat{\mbox{\boldmath$L$}^{\prime }}}
{\partial A^{\prime}{}_{\mu ,\nu}}\right)d{\sigma}^{\prime}{\hspace{0.01mm}}_{\mu} \;,
\end{equation}
with the Lagrangian density $\hat{\mbox{\boldmath$L$}^{\prime }}$ defined with transformed field variables, is evaluated as  
\begin{equation}
\label{eq:charge-trans-eq-hat}
\hat{C}^{\prime}=\hat{C}=q\; ,
\end{equation}
without imposing any additional condition.

To summarize, the G. E. P. is satisfied if the conditions in Eq. (\ref{eq:L-J-hat-cond}) are satisfied. 
The asymptotic behavior of the transformed dual components of 
$\hat{b}^{\prime k}{}_{\mu}
\stackrel{\mbox{\scriptsize def}}{=}
H^{k}{}_{l}(\partial x^{\nu }/\partial x^{\prime\mu})
b^{l}{}_{\nu }$ and of the transformed vector potential 
$\hat{A}^{\prime}{}_{\mu}
\stackrel{\mbox{\scriptsize def}}{=}(\partial x^{\nu }/\partial x^{\prime\mu })A_{\nu }$ are 
easily determined from Eqs. (\ref{eq:sl2ctr}), (\ref{eq:omega1}), 
(\ref{eq:inverse-coordinate}) and (\ref{eq:L-J-hat-cond}) as 
\begin{eqnarray}
\label{eq:b-prime-hat-asympt}
\hat{b}^{\prime k }{}_{\mu }
&=&\Lambda^{k}{}_{l}(C^{-1})^{\nu }{}_{\mu }{}^{(0)}b^{l}{}_{\nu }+
O^{k}{}_{\mu }(1/r^{p})\; , \; \; \; \left( p>\frac{1}{2}\right )\; \nonumber \\
\label{eq:A-prime-hat-asympt}
\hat{A}^{\prime}{}_{\mu}&=&
(C^{-1})^{\nu }{}_{\mu }A_{\nu }+d^{\nu }{}_{\mu }A_{\nu }
=O_{\mu }(1/r)\; .
\end{eqnarray}
\section{Summary and discussion}
\label{sec:summary}

In extended new general relativity (E. N. G. R.), we have examined exact charged axi-symmetric solutions of the 
gravitational and electromagnetic field equations in vacuum from 
the point of view of the equivalence principle.

In this theory, the generators depend on the choice of the set of
independent field variables. In \S 3, we examined the solution represented by Eqs.~(\ref{eq:v-fields}) and (\ref{eq:em-potential})
for the case in which $\{\psi^{k},A^{k}{}_{\mu },A_{\mu }\}$ is employed as the set of independent field variables and for the case in which $\{\psi^{k},b^{k}{}_{\mu },A_{\mu }\}$ is employed as the set of independent variables. We have shown the following:
\begin{description}
\item[(A)]$\;$For the case in which $\{\psi^{k},A^{k}{}_{\mu},A_{\mu }\}$ is
employed as the set of independent field variables, the total energy-momentum, the total angular momentum and the total 
electric charge of the system are all given by generators of {\em internal} 
transformations. The canonical energy-momentum and the orbital angular momentum
vanish trivially.
\item[(B)]$\;$For the case in which $\{\psi^{k},b^{k}{}_{\mu},A_{\mu }\}$ is 
employed as the set of independent field variables, we have following: (1)$\; $The total energy-momentum is given by the generator $M^{c}{}_{\mu }$
of the coordinate translations, and the generator $\hat{M}_{k}$ of the 
internal translations vanishes identically. (2)$\; $The total angular 
momentum is given by the sum $\hat{J}_{kl}\stackrel{\mbox{\scriptsize def}}{=}
\hat{S}_{kl}+{}^{(0)}b^{\mu}{}_{k}{}^{(0)}b^{\nu}{}_{l}\hat{L}{}_{[\mu \nu]}$ 
of the generator $\hat{S}_{kl}$ of the internal Lorentz transformations and of the generator $\hat{L}{}_{[\mu \nu ]}$ of the coordinate Lorentz transformations. (3)$\; $The total charge is given by the generators of the internal 
$U(1)$ transformations. 
\item[(A$\cap$B)]$\; $For both cases, the total energy-momentum, the total angular momentum and the total charge of the system are identical to the corresponding active quantities of a central gravitating body. \end{description}

The total mass, which is equal to the total energy divided by the square of the velocity of light, can be regarded as the inertial mass of the system. Thus, the results mentioned above include the equality of the inertial mass and the active gravitational mass, which implies 
that the equivalence principle is satisfied by the solution represented by Eqs.~(\ref{eq:v-fields}) and (\ref{eq:em-potential}). In consideration of this, we have introduced the notion of a 
generalized equivalence principle (G. E. P.), as stated at the beginning 
of \S 4. Solutions obtained by applying 
\{the transformations (\ref{eq:sl2ctr}) and (\ref{eq:inttr}) satisfying the 
conditions (\ref{eq:restricted-local-Lorentz}) and 
(\ref{eq:spherical})$\}\otimes \{$the coordinate transformation 
(\ref{eq:coordinate})\} to the original solution have been examined from the 
point of view of the G. E. P. The following results have been obtained. 
\begin{description}
\item[(A$^{\prime}$)]$\,$For the case in which 
$\{\psi^{k},A^{k}{}_{\mu},A_{\mu }\}$ is employed as the set of independent field variables, the G. E. P. is satisfied by solutions if the conditions in
Eq. (\ref{eq:parameter-condition}) are satisfied. 
\item[(B$^{\prime}$)]$\,$For the case in which $\{\psi^{k},b^{k}{}_{\mu},
A_{\mu }\}$ is employed as the set of independent field variables, the G. E. P. is satisfied if the conditions in Eq.~(\ref{eq:L-J-hat-cond}) are satisfied. \\
\end{description}

We would like to add several comments:
\begin{description}
\item[{[1]}]$\;$For the internal transformation (\ref{eq:sl2ctr}), the condition (\ref{eq:parameter-condition}) gives a stronger condition 
than does the condition (\ref{eq:L-J-hat-cond}). For the coordinate transformation 
(\ref{eq:coordinate}), no constraint is imposed by the former, while the latter gives 
the restrictions $u>1,p+u>2$.
\item[{[2]}]$\;$For the case in which $\{\psi^{k},A^{k}{}_{\mu},A_{\mu }\}$ is employed as the set of independent field variables, the G. E. P. {\em
is satisfied even by solutions which approach constant values very slowly 
at spatial infinity}, as is seen from Eq. (\ref{eq:b-A-prime-asympt}).  

This is a direct consequence of the following:
\begin{description}
\item[(a)]$\;${\em The functions $\mbox{\boldmath$F$}_{k}{}^{\mu\nu}$ and ${\mbox{\boldmath$F$}}^{\mu \nu }=\sqrt{-g}F^{\mu \nu }$ both describe tensor densities}, and hence the total energy-momentum $M_{k}$, total angular momentum
$S_{kl}$ and the total charge $C$ are all independent of the coordinate systems employed. 
\item[(b)]$\;$The canonical energy-momentum $M^{c}{}_{\mu }$ and 
the \lq \lq extended orbital angular momentum" $L_{\mu }{}^{\nu }$, 
which are the generators of coordinate transformations, obey the regular transformation rules (\ref{eq:CEM-trans}) and (\ref{eq:L-trans})
under the coordinate transformation 
(\ref{eq:coordinate}) with {\em arbitrary positive} $u$ if the 
condition (\ref{eq:L-cond}) is satisfied.
\end{description}  
Note that $b^{k\prime }{}_{\mu \prime }$ as given by 
Eq.~(\ref{eq:b-A-prime-asympt}) approaches a constant value much more slowly than the vierbein components required in the general situation to give reasonable forms of energy-momentum and angular momentum.\cite{Kawai-Toma}
\item[{[3]}]$\;$
In the case in which $\{\psi^{k},b^{k}{}_{\mu},A_{\mu }\}$ is employed as the set of independent field variables, the G.E.P. for the total energy-momentum and for the total angular momentum is established when the vierbeins 
have asymptotic property as indicated by the first of Eq. (\ref{eq:A-prime-hat-asympt}).
This is consistent with the results \cite{Shirafuji-Nashed-Hayashi} on the equivalence principle for the energy in new general relativity (N. G. R.).\footnote{Note that E. N. G. R. is reduced to N. G. R. if fields with nonvanishing $P_{k}$ are not present and if the set 
$\{\psi^{k}, b^{k}{}_{\mu },A_{\mu },\phi^{A},\phi^{*A}\}$ is employed as the set of independent field variables.}   
\end{description}

The preceding results, together with those in Ref.\cite{Kawai-Toma}, show that the choice $\{\psi^{k}, A^{k}{}_{\mu },A_{\mu }\}$ as the set of independent field variables is preferential to the choice 
$\{\psi^{k}, b^{k}{}_{\mu },A_{\mu }\}$. This is quite natural, because 
the fields $\psi^{k}, A^{k}{}_{\mu }$ and $A_{\mu }$ are 
the fundamental objects and $b^{k}{}_{\mu }$ is a composite of $\psi^{k}$ 
and $A^{k}{}_{\mu }$.
\appendix \section{Asymptotic Forms of ${\mbox{\boldmath$F$}}^{(1)}{}_{k}{}^{\mu \nu }\; ,
V^{mn}{}_{\lambda}$ and $W^{\mu \nu \lambda }{}_{klm}$ for Large $r$}
In this appendix, we give the asymptotic forms of ${\mbox{\boldmath$F$}}^{(1)}{}_{k}{}^{\mu \nu }\; ,
V^{mn}{}_{\lambda}$ and $W^{\mu \nu \lambda }{}_{klm}$
for large $r$, which are useful for the calculations in  \S \S 3 and 4.\\
{}\\
${\mbox{\boldmath$F$}}^{(1)}{}_{k}{}^{\mu \nu }$:
\begin{eqnarray}
\label{eq:asympt-F}
{\mbox{\boldmath$F$}}^{(1)}{}_{(0)}{}^{0\alpha }&=&-\frac{1}{\kappa }\left(a-\frac{Q^{2}}{r}\right)\frac{n^{\alpha }}{r^{2}}+\frac{ah}{2\kappa r^{3}}
\varepsilon^{\alpha}{}_{\beta 3}n^{\beta }
+O^{\alpha }\left(\frac{1}{r^{4}}\right)\; ,\nonumber \\
{\mbox{\boldmath$F$}}^{(1)}{}_{a}{}^{0\alpha }&=&-\frac{1}{\kappa r^{2}}\left(a-
\frac{3Q^{2}}{2r}\right)n_{a}n^{\alpha }
-\frac{Q^{2}}{2\kappa r^{3}}\delta_{a}{}^{\alpha }\nonumber \\
& &-\frac{ah}{2\kappa r^{3}}\left(\delta^{\alpha \beta }
-3n^{\alpha }n^{\beta }\right)\varepsilon_{a\beta 3}
+O_{a}{}^{\alpha }\left(\frac{1}{r^{4}}\right)\; ,\nonumber \\
{\mbox{\boldmath$F$}}^{(1)}{}_{(0)}{}^{\alpha \beta}&=&
\frac{3ah}{2\kappa r^{3}}
\left(n^{\alpha }\varepsilon^{\beta}{}_{\gamma 3}-n^{\beta }
\varepsilon^{\alpha}{}_{\gamma 3}\right)n^{\gamma }
+\frac{ah}{\kappa r^{3}}\varepsilon^{\alpha \beta}{}_{3}
+O^{[\alpha \beta ]}\left(\frac{1}{r^{4}}\right)\; ,\nonumber \\
{\mbox{\boldmath$F$}}^{(1)}{}_{a}{}^{\alpha \beta}&=&
-\frac{ah}{2\kappa r^{3}}
\left(\varepsilon_{a}{}^{\alpha }{}_{3}n^{\beta }
-\varepsilon_{a}{}^{\beta}{}_{3}n^{\alpha }\right)
+\frac{2ah}{\kappa r^{3}}
\left(n^{\alpha }\varepsilon^{\beta}{}_{\gamma 3}-n^{\beta }
\varepsilon^{\alpha}{}_{\gamma 3}\right)n_{a}n^{\gamma } \nonumber \\
& &+\frac{ah}{\kappa r^{3}}\varepsilon^{\alpha \beta}{}_{3}n_{a}
-\frac{Q^{2}}{2\kappa r^{3}}\left(\delta_{a}{}^{\alpha }n^{\beta }
-\delta_{a}{}^{\beta }n^{\alpha }\right)
+O_{a}{}^{[\alpha \beta]}\left(\frac{1}{r^{4}}\right)\; .
\end{eqnarray}
{}\\
$V^{mn}{}_{\lambda}$:
\begin{eqnarray}
\label{eq:asympt-V}
V^{(0)a}{}_{0}&=&-n^{a}\dot{\Xi}-a\frac{n^{a}}{r}\dot{\Pi}
-\frac{n^{a}}{2}\Xi^{2}(\dot{\Xi}-\frac{a}{r}\dot{\Pi})
+a\frac{n^{a}}{r}\Xi \Pi\dot{\Xi}\nonumber \\
& &+O^{a}\left(\frac{1}{r^{s+1}}\right)
+O^{a}\left(\frac{1}{r^{2p+2}}\right)
+o^{a}\left(\frac{1}{r^{3}}\right)\; ,\nonumber \\
V^{(0)a}{}_{\alpha }&=&-\frac{1}{r}\left(\delta^{a}{}_{\alpha }-n^{a}n_{\alpha }
\right)\Xi-n^{a}n_{\alpha }\Xi^{\prime }
-\frac{a}{r^{2}}\left(\delta^{a}{}_{\alpha}
-2n^{a}n_{\alpha }\right)\Pi \nonumber \\
& &-\frac{an^{a}n_{\alpha }}{r}\Pi^{\prime }+
\frac{n^{a}n_{\alpha }}{2}\Xi^{2}\Xi^{\prime }+
O^{a}{}_{\alpha }\left(\frac{1}{r^{s+1}}\right)\nonumber \\
& &+O^{a}{}_{\alpha }\left(\frac{1}{r^{2p+2}}\right)
+o^{a}{}_{\alpha }\left(\frac{1}{r^{3}}\right)\; ,\nonumber \\
V^{ab}{}_{0}&=&-\frac{a}{r}n^{a}n^{b}(\Xi \dot{\Pi}+\dot{\Xi}\Pi)
-a^{2}\frac{n^{a}n^{b}}{r^{2}}\Pi \dot{\Pi}+\frac{1}{2}
n^{a}n^{b}\Xi^{3}\dot{\Xi}\nonumber \\ 
& &+O^{ab}\left(\frac{1}{r^{s+1}}\right)
+O^{ab}\left(\frac{1}{r^{2p+2}}\right)
+o^{ab}\left(\frac{1}{r^{3}}\right)\; ,\nonumber \\
V^{ab}{}_{\alpha }&=&-\frac{1}{r}
(n^{a}\delta^{b}{}_{\alpha }-n^{b}\delta^{a}{}_{\alpha })
\left(1-\frac{1}{4}\Xi^{2}\right)\Xi^{2}
+O^{ab}{}_{\alpha }\left(\frac{1}{r^{s+1}}\right)\nonumber \\
& &+O^{ab}{}_{\alpha }\left(\frac{1}{r^{2p+2}}\right)
+o^{ab}{}_{\alpha }\left(\frac{1}{r^{3}}\right)
\end{eqnarray}
with\footnote{Here, the prime and dot represent derivatives with respect
to $r$ and $x^{0}$, respectively. For example, 
$A^{\prime }\stackrel{\mbox{\scriptsize def}}{=}\partial A/\partial r \;$ and
$\dot{A}\stackrel{\mbox{\scriptsize def}}{=}\partial A/\partial x^{0}$.}
\begin{equation}
\label{eq:Xi-Pi}
\Xi \stackrel{\mbox{\scriptsize def}}{=}A^{\prime }+\dot{B}\; ,\; \; \; 
\Pi \stackrel{\mbox{\scriptsize def}}{=}\dot{A}+\dot{B}-A^{\prime }-B^{\prime }
+\frac{1}{r}(A+B)\; ,
\end{equation}
and $o^{a}{}_{\alpha }(1/r^{3})$, for example, denotes a term such that
$\lim_{r\rightarrow \infty}r^{3}o^{a}{}_{\alpha }(1/r^{3})=0$.\\
{}\\
$W^{\mu \nu \lambda}{}_{klm}$:
\begin{eqnarray}
\label{eq:asympt-W}
W^{0\alpha \beta }{}_{(0)ab}&=&\frac{1}{2}\left(\delta^{\alpha }{}_{a}
\delta^{\beta}{}_{b}-\delta^{\alpha}{}_{b}\delta^{\beta}{}_{a}\right)
+\frac{a}{4r}\left\{\left(\delta^{\alpha}{}_{a}\delta^{\beta}{}_{b}
-\delta^{\alpha}{}_{b}\delta^{\beta}{}_{a}\right)
+\left(n_{a}\delta^{\alpha }{}_{b}-n_{b}\delta^{\alpha}{}_{a}\right)n^{\beta}\right. \nonumber \\
& &\left. +\left(n_{b}\delta^{\beta }{}_{a}-n_{a}\delta^{\beta}{}_{b}\right)
n^{\alpha}\right\}+O^{[\alpha \beta ]}{}_{[ab]}\left(\frac{1}{r^{2}}
\right)\; ,\nonumber \\
W^{0\alpha \beta }{}_{abc}&=&\frac{a}{4r}
\left\{\left(\delta^{\alpha}{}_{a}\delta^{\beta}{}_{b}
-\delta^{\alpha}{}_{b}\delta^{\beta}{}_{a}\right)n_{c}
+\left(\delta^{\alpha}{}_{b}\delta^{\beta}{}_{c}
-\delta^{\alpha}{}_{c}\delta^{\beta}{}_{b}\right)n_{a}\right .\nonumber \\
&&+\left .\left(\delta^{\alpha}{}_{c}\delta^{\beta}{}_{a}
-\delta^{\alpha}{}_{a}\delta^{\beta}{}_{c}\right)n_{b}\right\}
+O^{[\alpha \beta ]}{}_{[abc]}\left(\frac{1}{r^{2}}\right)
\; ,\nonumber \\
W^{\alpha \beta \gamma}{}_{(0)ab}&=&
-\frac{a}{4r}
\left\{\left(\delta^{\alpha}{}_{a}\delta^{\beta }{}_{b}-\delta^{\alpha }{}_{b}\delta^{\beta }{}_{a}\right)n^{\gamma}
+\left(\delta^{\beta}{}_{a}\delta^{\gamma}{}_{b}-\delta^{\beta}{}_{b}
\delta^{\gamma}{}_{a}\right)n^{\alpha}\right . .\nonumber \\
&&+\left .\left(\delta^{\gamma }{}_{a}\delta^{\alpha}{}_{b}-\delta^{\gamma}{}_{b}
\delta^{\alpha}{}_{a}\right)n^{\beta}\right\}
+O^{[\alpha \beta \gamma]}{}_{[ab]}\left(\frac{1}{r^{2}}\right)
\; ,\nonumber \\
W^{\alpha \beta \gamma}{}_{abc}&=&
\frac{1}{2}\left\{\left(\delta^{\alpha }{}_{a}\delta^{\beta}{}_{b}-
\delta^{\beta}{}_{a}\delta^{\alpha}{}_{b}\right)\delta^{\gamma}{}_{c}
+\left(\delta^{\alpha }{}_{b}\delta^{\beta}{}_{c}-
\delta^{\beta}{}_{b}\delta^{\alpha}{}_{c}\right)\delta^{\gamma}{}_{a}\right .\nonumber\\
&&+\left .\left(\delta^{\alpha }{}_{c}\delta^{\beta}{}_{a}-
\delta^{\beta}{}_{c}\delta^{\alpha}{}_{a}\right)\delta^{\gamma}{}_{b}\right\}
\nonumber \\
& &-\frac{a}{4r}\left\{\left(\delta^{\alpha}{}_{a}
\delta^{\beta}{}_{b}-\delta^{\beta }{}_{a}\delta^{\alpha}{}_{b}\right)n_{c}
+\left(\delta^{\alpha}{}_{b}\delta^{\beta}{}_{c}-\delta^{\beta }{}_{b}
\delta^{\alpha}{}_{c}\right)n_{a}\right .\nonumber \\
&&+\left .\left(\delta^{\alpha}{}_{c}\delta^{\beta}{}_{a}-\delta^{\beta }{}_{c}
\delta^{\alpha}{}_{a}\right)n_{b}\right\}n^{\gamma}\nonumber \\
& &
-\frac{a}{4r}\left[\left\{\left(\delta^{\alpha}{}_{a}n_{b}
-\delta^{\alpha }{}_{b}n_{a}\right)n^{\beta }
-\left(\delta^{\beta}{}_{a}n_{b}-\delta^{\beta}{}_{b}n_{a}\right)n^{\alpha}
\right\}\delta^{\gamma}{}_{c}\right. \nonumber \\
& &\left.+\left\{\left(\delta^{\alpha}{}_{b}n_{c}-\delta^{\alpha }{}_{c}n_{b}\right)n^{\beta }
-\left(\delta^{\beta}{}_{b}n_{c}-\delta^{\beta}{}_{c}n_{b}\right)n^{\alpha}
\right\}\delta^{\gamma}{}_{a}\right.\nonumber \\
& &\left.+\left\{\left(\delta^{\alpha}{}_{c}n_{a}-\delta^{\alpha }{}_{a}n_{c}
\right)n^{\beta }
-\left(\delta^{\beta}{}_{c}n_{a}-\delta^{\beta}{}_{a}n_{c}\right)n^{\alpha}
\right\}\delta^{\gamma}{}_{b}\right]\nonumber \\
&&+O^{[\alpha \beta \gamma]}{}_{[abc]}\left(\frac{1}{r^{2}}\right)\;
.
\end{eqnarray}

All the other components, which are not listed above, are equal to zero
or are obtainable from the listed ones by permutations.

\end{document}